\documentclass[twocolumn,tighten]{aastex63}

\usepackage{amsmath}
\usepackage{amsbsy}
\usepackage{rotating}

\shorttitle{The S-Web Origin of Solar Wind Composition Enhancement}
\shortauthors{Lynch et al.}
\hypersetup{linkcolor=magenta,citecolor=blue,filecolor=cyan,urlcolor=purple}

\begin{document}


\title{The S-Web Origin of Composition Enhancement in the Slow-to-Moderate Speed Solar Wind}


\correspondingauthor{Benjamin~J.~Lynch}
\email{blynch@berkeley.edu}

\author[0000-0001-6886-855X]{B.~J.~Lynch}
\affil{Space Sciences Laboratory, University of California--Berkeley, Berkeley, CA 94720, USA}

\author[0000-0003-1692-1704]{N.~M.~Viall}
\affil{NASA Goddard Space Flight Center, Greenbelt, MD 20771, USA}

\author[0000-0003-1380-8722]{A.~K.~Higginson}
\affil{NASA Goddard Space Flight Center, Greenbelt, MD 20771, USA}

\author[0000-0002-5975-7476]{L.~Zhao}
\affil{Department of Climate and Space Sciences and Engineering, University of Michigan, Ann Arbor, MI 48109, USA}

\author[0000-0003-1611-227X]{S.~T.~Lepri} 
\affil{Department of Climate and Space Sciences and Engineering, University of Michigan, Ann Arbor, MI 48109, USA}

\author[0000-0003-4043-616X]{X.~Sun}
\affil{Institute for Astronomy, University of Hawaii--Manoa, Pukalani, HI 96768, USA}

%

\begin{abstract}

Connecting the solar wind observed throughout the heliosphere to its origins in the solar corona is one of the central aims of heliophysics. The variability in the magnetic field, bulk plasma, and heavy ion composition properties of the slow wind are thought to result from magnetic reconnection processes in the solar corona. We identify regions of enhanced variability and composition in the solar wind from 2003 April 15 to May 13 (Carrington Rotation 2002), observed by the \emph{Wind} and \emph{Advanced Composition Explorer} spacecraft, and demonstrate their relationship to the Separatrix--Web (S-Web) structures describing the corona's large-scale magnetic topology. There are four pseudostreamer (PS) wind intervals and two helmet streamer (HS) heliospheric current sheet/plasma sheet crossings (and an ICME) which all exhibit enhanced alpha-to-proton ratios and/or elevated ionic charge states of carbon, oxygen, and iron. We apply the magnetic helicity--partial variance of increments ($H_m$--PVI) procedure to identify coherent magnetic structures and quantify their properties during each interval. The mean duration of these structures are $\sim$1~hr in both the HS and PS wind. We find a modest enhancement above the power-law fit to the PVI waiting time distribution in the HS-associated wind at the 1.5--2~hr timescales that is absent from the PS intervals. We discuss our results in context of previous observations of the $\sim$90~min periodic density structures in the slow solar wind, further development of the dynamic S-Web model, and future \emph{Parker Solar Probe} and \emph{Solar Orbiter} joint observational campaigns.

\end{abstract}

\keywords{solar wind ---  Sun: heliosphere --- 
          Sun: corona --- Sun: magnetic fields --- Sun: solar-terrestrial relations}

\section{Introduction}
\label{sec:intro}

The global magnetic geometry of the solar corona directly determines the structure of the solar wind outflow \citep[e.g.][]{Zirker1977, Axford1999, Antiochos2007, Antiochos2011, Cranmer2012}. Decades of in-situ observations have shown that the heliospheric structure and solar wind properties reflect the coronal magnetic structure of its origin \citep{Zurbuchen2007, ZhaoL2014c}. During solar minimum, polar coronal holes are correlated with fast, tenuous solar wind \citep{Geiss1995b, McComas2002}, while the helmet streamer (HS) belt and the heliospheric current sheet (HCS) are associated with slower, denser, and more variable solar wind \citep{Gosling1997, McComas1998a, Zurbuchen2002, ZhaoL2009}. During solar maximum, the helmet streamer belt is highly warped and pseudostreamer (PS) coronal structures often make a significant contribution to the solar wind in the ecliptic plane \citep{Riley2012}.

Whereas the large-scale closed flux system of the HS belt separates open fields of opposite polarity, thus giving rise to the HCS, coronal PS's (sometimes called unipolar streamers) are closed-flux regions surrounded by open fields of a single polarity \citep[e.g.][]{WangYM2007a, Titov2012, Rachmeler2014, WangYM2012a, WangYM2019a, Mason2021}. Solar wind originating from coronal PS's tends to be more similar to the dense, variable HS slow wind than to the fast wind from coronal holes \citep{Crooker2012a}, but observations have established the existence of a continuum of states between the nominal fast and slow wind rather than a well-separated bimodal distribution \citep[e.g.][]{Stakhiv2015, Stakhiv2016}. 

Connecting the solar wind to its source region of origin has become one of the central aims of heliophysics in order to test and constrain different theories of solar wind formation \citep{Viall2020}. Additionally, accurate space weather prediction requires an understanding of the different solar wind streams in the heliosphere and where they were formed, e.g. mesoscale structures are known to drive magnetospheric dynamics \citep{Viall2021}. Therefore, establishing this solar--heliospheric connection is one of the fundamental science objectives of the \emph{Parker Solar Probe} \citep[PSP;][]{Fox2016} and \emph{Solar Orbiter} \citep[][]{Mueller2020} missions.

White-light coronagraph and heliospheric imaging data have shown that the solar wind originating from the helmet streamer stalks includes a continual, intermittent outflow of intensity enhancements, called ``streamer blobs," that trace the bulk outflow of the slow solar wind \citep{Sheeley1997, Sheeley1999, Sheeley2009, Rouillard2010a, Rouillard2010b,SanchezDiaz2017a}. While the basic theory of steady-state, slow solar wind from the vicinity of coronal streamers and pseudostreamers is well-established \citep[e.g.][and references therein]{Arge2000, Lepri2008, Riley2012}, this steady-state picture is difficult to reconcile with the observed slow wind variability in both remote-sensing and in-situ observations that likely require a time-varying magnetic reconnection component.

Demonstrating another example of solar wind variability, \citet{Kepko2020} analyzed 25 years of solar wind data, expanding on the initial study of \citet{Viall2008}, finding that intermittent periodic density structures that range in size from 70--900~Mm are a ubiquitous feature of the slow solar wind, occurring a majority ($\gtrsim 60\%$) of the time. Furthermore, \citet{Viall2010} and \citet{Viall2015} examined the \emph{Solar Terrestrial Relations Observatory} \citep[STEREO;][]{kaiser2008} SECCHI \citep{HowardR2008} HI1 and COR2 white-light imaging data and showed there were clear signatures of $\sim$90~min variability in the intensity variations of coronal streamer outflow, confirming that many of the periodic density structures are the result of solar wind formation processes. 

In the in-situ slow solar wind, especially near the HCS, magnetic structures with timescales of several hours have been identified and linked to magnetic reconnection \citep{Crooker1996hps, Crooker2004, Suess2009}. High-cadence composition data have revealed the presence of cyclic 0.5--3 hour solar wind structures with signatures in helium, oxygen and carbon densities, and heavy ion charge states \citep{Viall2009b, Kepko2016}. In-situ elemental and ionic composition measurements are routinely used as proxies for solar wind formation processes and the ``freeze-in'' coronal electron temperatures in the low-to-middle corona; when the characteristic bulk solar wind expansion timescale exceeds the ionization and recombination timescales of various ion species, the ionic charge states can be considered frozen-in to the solar wind outflow \citep[e.g.][]{Hundhausen1968, Owocki1983, Ko1997, Landi2012a, Landi2012b, Landi2015}. In fact, \citet{Kepko2016} showed that these density and compositional variations also often correspond to regions of coherent magnetic field signatures and periods of bidirectional electron streaming, suggestive of a succession of small magnetic flux ropes or flux rope-like periods. There is some preliminary indication that in-situ small flux ropes can be coincident with periods of enhanced ionic composition \citep{Foullon2011, Feng2015a, Yu2016, Kepko2016}.

The Separatrix-Web (S-Web) model for the origin of slow solar wind \citep{Antiochos2011} is based on the magnetic geometry of the solar corona and predicts that the topological separatrix surfaces of the magnetic field are regions where interchange reconnection---the mechanism for releasing closed-flux coronal plasma onto adjacent open field lines---is most likely to occur. 

The dynamic S-Web model extends previous observational and theoretical considerations of reconnection at coronal hole boundaries \citep{Madjarska2004, Edmondson2009, Edmondson2010a, Linker2011, Rappazzo2012, Brooks2015, Pontin2015, Scott2021} and solar wind outflows at the periphery of active regions \citep[e.g.][]{Sakao2007, Harra2008, Baker2009, Brooks2011, Edwards2016}, and aims to address a number of outstanding issues related to the slow solar wind, including its larger-than-expected latitudinal extent \citep{Crooker2012a} and the reconnection component seemingly required by the variability of the in-situ measurements of slow wind plasma, field, and composition \citep[e.g.][]{Viall2009b, ZhaoL2009, ZhaoL2014c, ZhaoL2017a, Lepri2013, Lepri2014IAU, Kepko2016,SanchezDiaz2017b,SanchezDiaz2019,DiMatteo2019,Reville2022}.

\begin{figure*}[!t]
	\centerline{ \includegraphics[width=1.0\textwidth]{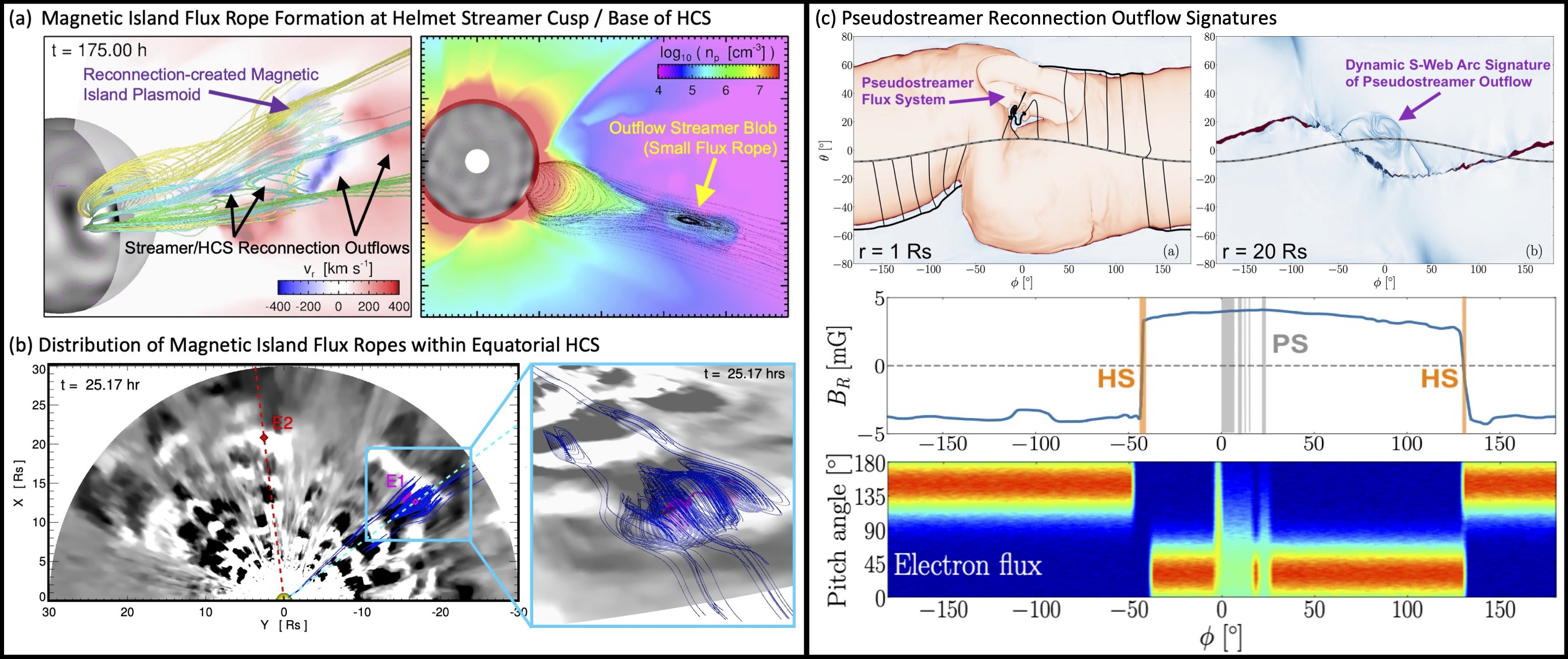} }
	\caption{Reconnection mechanisms for generating 
           intermittent outflow of dense, closed-field plasma in the slow-to-moderate speed solar wind from helmet streamers and pseudostreamers. (a) ARMS simulation of HS blob pinch-off 
           reconnection \citep[adapted from][]{Lynch2020} and (b) the small flux rope/reconnection plasmoid structures of the heliospheric current sheet \citep[adapted from][]{Higginson2018}. (c) ARMS simulation of interchange reconnection outflow from a pseudostreamer and a synthetic proxy for suprathermal electron pitch angle based on magnetic connectivity \citep[adapted from][]{Aslanyan2022}.}
	\label{fig1}
\end{figure*}

\citet{Higginson2017a,Higginson2017b} presented simulation results showing that interchange magnetic reconnection is ubiquitous and most likely responsible for releasing much of the slow solar wind, in particular along S-Web topological features. Since that work, there have been a number of significant developments in the modeling reconnection-generated slow solar wind structure and the interchange reconnection processes associated with dynamic S-Web outflows, summarized in Figure~\ref{fig1}. Figure~\ref{fig1}(a) presents the 3D structure of the pinch-off reconnection that forms streamer blob flux rope/plasmoids in the simulation by  \citet{Lynch2020}. These simulation results showed qualitative agreement with both the morphology and the kinematics of coronal inflows and streamer blob outflows in synthetic white light coronagraph observations, as have other recent modeling efforts \citep[e.g.][]{Reville2020}. Figure~\ref{fig1}(b) presents simulation results from \citet{Higginson2018} who showed that the continual formation of flux rope/plasmoid structures essentially filled the entire heliospheric current sheet. Figure~\ref{fig1}(c) shows the simulation results by \citet{Aslanyan2022} in which they examined interchange reconnection occurring in a 3D pseudostreamer configuration and developed a synthetic suprathermal electron pitch angle proxy based on the simulation's instantaneous magnetic connectivity.

Previously, \citet{ZhaoL2017a} have used solar wind data from the \emph{Advanced Composition Explorer} \citep[ACE;][]{Stone1998} during CR~2002 to develop a source region classification scheme based on heliospheric back-mapping and PFSS modeling of observer-connected magnetic field lines and the pixel brightness in synoptic maps of EUV 195\AA\ emission in the vicinity of the field line foot point. Applying their EUV brightness-based source region classifications (`Coronal Hole', `Coronal Hole Boundary', `Quiet Sun', `Active Region Boundary', `Active Region', and `Helmet Streamer') to in-situ data from 1998--2011 resulted in a statistical ordering of the distributions of O$^{7+}$/O$^{6+}$ by distance from coronal holes, representing a relatively smooth increase in some combination of coronal electron temperature, mass density, and/or outflow velocities.

A number of other solar wind classification schemes have been developed to identify specific solar wind ``types'' for the purpose of trying to uncover the physical relationships between different plasma, field, and composition signatures \emph{within} and \emph{between} different solar wind types (which are generally a proxy for coronal source region classifications). For example, \citet{Xu2015} constructed a ``four-plasma'' classification scheme based, in part, on the proton specific entropy, $S_p=T_p/n_p^{2/3}$, and showed this had a significant correlation with O$^{7+}$/O$^{6+}$, C$^{6+}$/C$^{5+}$ signatures and a relatively clear separation in the Alfv\'{e}n speed--specific entropy ($v_A$--$S_p$) space between their `Ejecta', `Coronal Hole', 'Streamer Belt', and 'Sector Reversal' (HCS/HPS crossing) types. \citet{Ko2018} examined the perpendicular velocity fluctuations ($\delta v_{T}$, $\delta v_{N}$ in RTN coordinates) and presented superposed epoch trends in HS and PS intervals (low-$\delta v$) for a variety of solar wind properties including magnetic field fluctations, Alfv\'{e}ncity, width of the suprathermal electron strahl, proton specific entropy $S_p$, helium abundance, the C, O, and Fe charge states, and Fe/O composition. \citet{Bloch2020} have investigated a couple of machine learning techniques to identify `Streamer Belt' and `Coronal Hole' solar wind type clusters in the $S_p$--O$^{7+}$/O$^{6+}$ parameter space from \emph{Ulysses} and ACE data. \citet{Roberts2020} have used $k$-means clustering based on a number of solar wind variables including O and Fe charge states and the Fe/O ratio which resulted in a mixture of some clearly separated solar wind types and some significantly overlapping solar wind types when visualized in the cross helicity ($\sigma_c$) and residual energy ($\sigma_r$) parameter space commonly used in turbulence studies.

In this paper, we extend the CR~2002 analysis of \citet{ZhaoL2017a} to the magnetic complexity of the source region and examine the relationship between measures of solar wind variability in plasma, field, and composition with the large-scale geometric S-Web configurations of the associated source regions. In Section~\ref{sec:int}, we present in-situ solar wind observations from the \emph{Wind} and ACE spacecraft during CR~2002 and define several slow-to-moderate speed intervals of enhanced variability in proton and alpha densities. We then show that each of these intervals correspond to enhancements in the ionic composition signatures of C, O, and Fe. In Section~\ref{sec:map}, we perform the heliospheric back-mapping procedure to map the in-situ time series at 1~au to Carrington longitude at the potential field source surface (PFSS) at $2.5\, R_\odot$ (\ref{sec:map:heliosphere}) and show these intervals of enhanced variability and composition map back to the S-Web topological structures associated with the helmet streamer belt and coronal pseudostreamers (\ref{sec:map:sweb}). In Section~\ref{sec:results}, we present the magnetic helicity--partial variance of increments ($H_m$--PVI) analysis during the enhanced variability intervals and quantify the similarities and differences between the helmet streamer (\ref{sec:results:hswind}) and pseudostreamer (\ref{sec:results:pswind}) slow wind, and perform some statistical analyses on these time series (\ref{sec:results:stats}). Finally, in Section~\ref{sec:disc}, we discuss the implications of our results for theory and modeling the origin of the slow solar wind and avenues for future progress with complementary PSP and \emph{Solar Orbiter} observations.

\begin{figure*}[!t]
    \centerline{\includegraphics[width=0.85\textwidth]{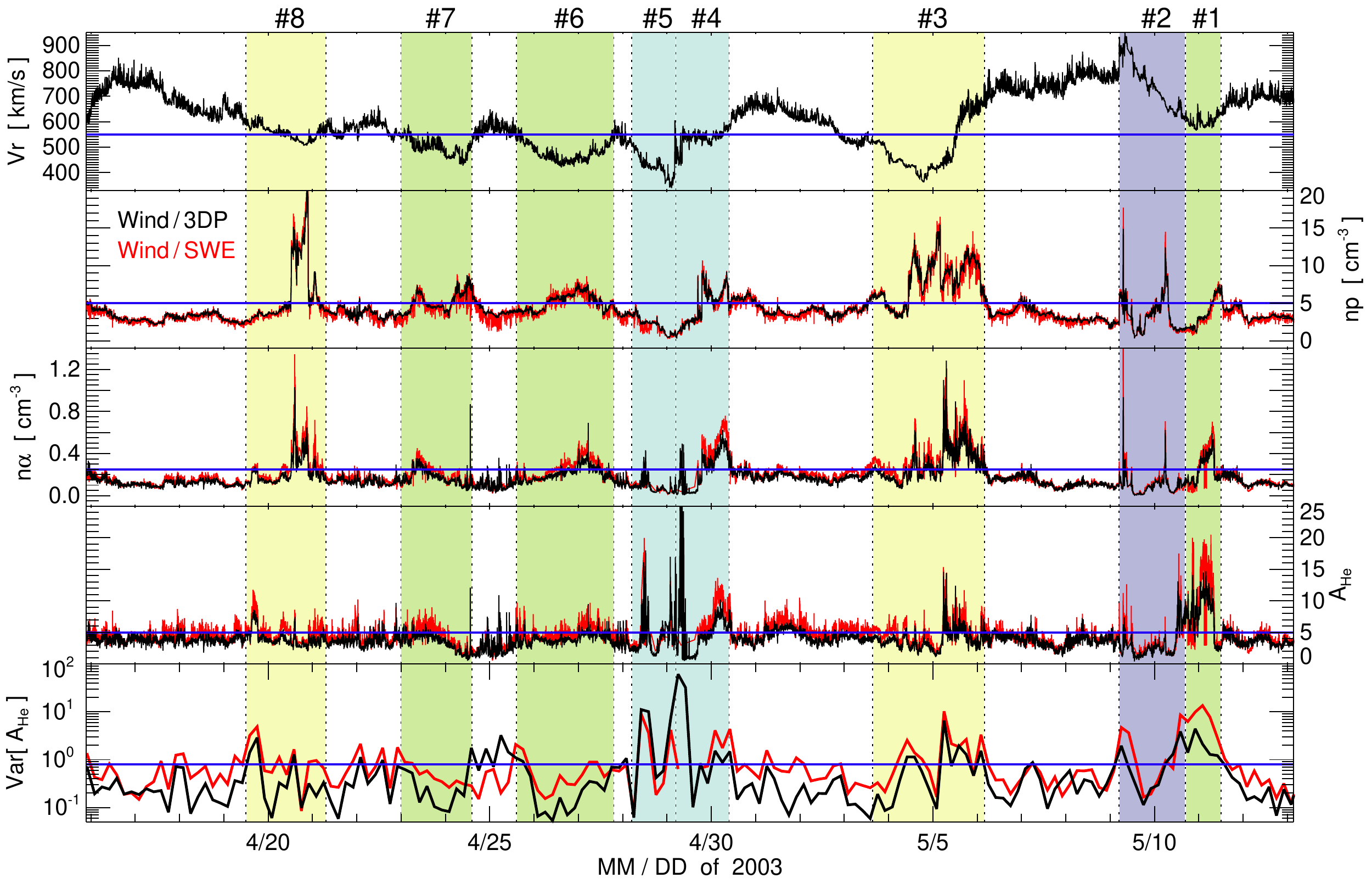} }
    \caption{In-situ solar wind data from the \emph{Wind} spacecraft during Carrington Rotation (CR) 2002. Plotted, from top-to-bottom, are the bulk radial velocity $V_r$, proton number density $n_p$, alpha number density $n_{\alpha}$, the ${\rm He}^{2+}/{\rm H}^{+}$ ratio, $A_{\rm He} \equiv n_\alpha/n_p \times 100$, and its variance $\sigma^2_{\alpha/p}$ in 6-hour bins. The black curves are \emph{Wind}/3DP 1~min data and the red curves are \emph{Wind}/SWE 97~s data. The eight intervals, labeled \#1--8 along the top axis, represent the different large-scale coronal source region classifications (HS---yellow, PS---green, teal, ICME---purple). The interval properties are summarized in Table~\ref{tab:int}. }
    \label{fig:int}
\end{figure*}

\begin{figure*}[!t]
    \centerline{ \includegraphics[width=0.85\textwidth]{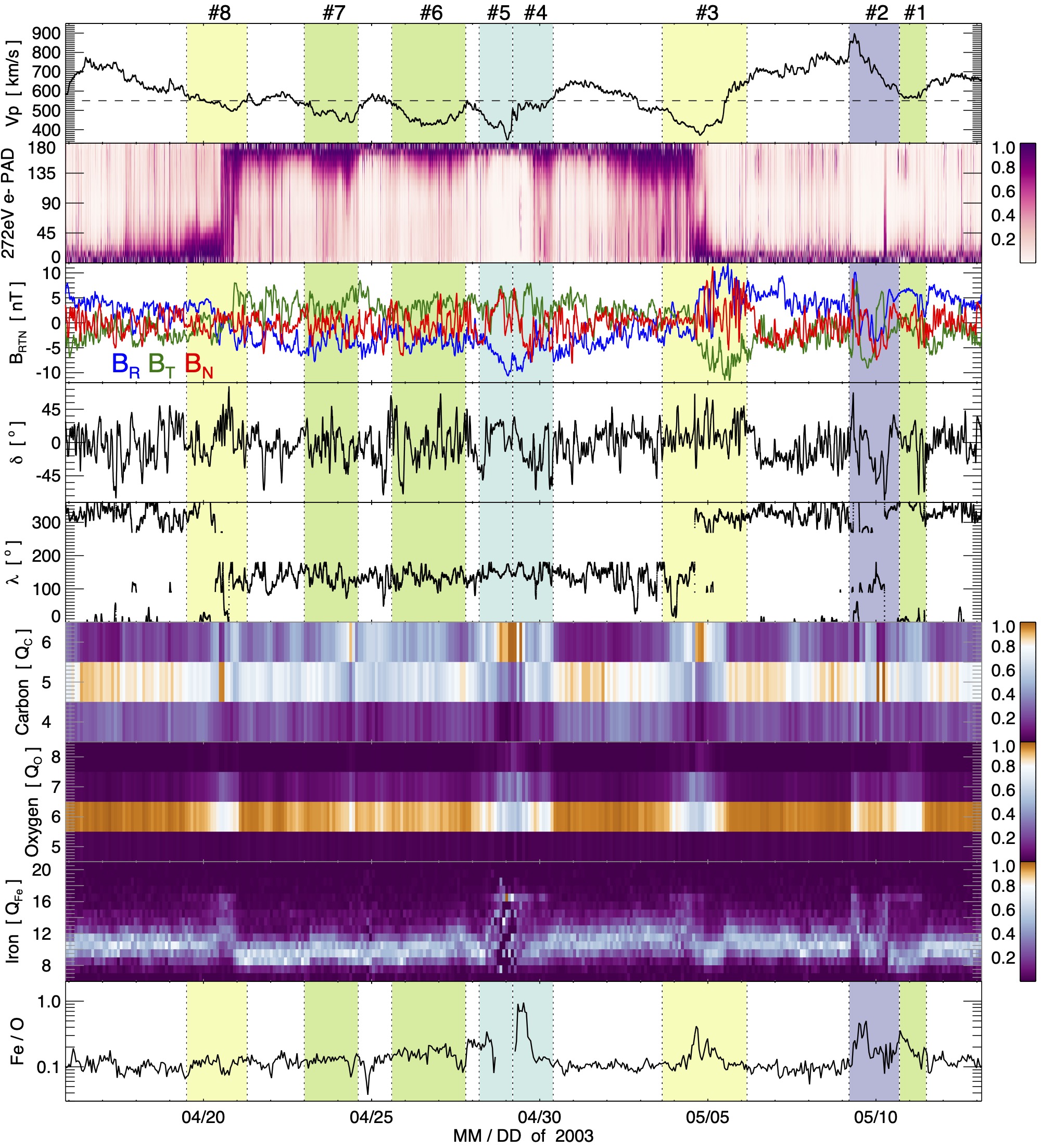} }
    \caption{Solar wind and ionic and elemental composition properties from ACE/SWEPAM, ACE/MAG, and ACE/SWICS for CR~2002. From top-to-bottom we plot: proton $V_r$, the 272~eV suprathermal electron pitch angle distribution (PAD), the magnetic field components of $\boldsymbol{B}_{\rm RTN}$, the magnetic field elevation and azimuthal angles ($\delta$, $\lambda$), the distribution of C$^{4-6+}$, O$^{5-8+}$, and Fe$^{6-20+}$, and the Fe/O ratio. The slow-to-moderate speed intervals from Figure~\ref{fig:int} and Table~\ref{tab:int} are also shown.}
    \label{fig:comp}
\end{figure*}

\section{Intervals of Enhanced Variability}
\label{sec:int}


\subsection{Proton Density and the Alpha-to-Proton Ratio}
\label{sec:int:wind}

The slow solar wind shows considerably more variation in proton and helium densities (and their relative abundance ratio) than in the fast wind. The mean alpha particle (He$^{2+}$) to proton (H$^{+}$) ratio $A_{\rm He} \equiv n_\alpha/n_p \times 100$ (or $\alpha/p$, interchangeably) in both the fast and slow solar wind are on the order of 3--5\% but the relative variation in the fast solar wind is $\sim$10\% while in the slow solar wind it can be as high as $\sim$40\% \citep{Gosling1997, Schwenn2006b}. Helium enhancements have long been associated with in-situ observations of CME material \citep[e.g.][]{Borrini1982, Richardson2004, Zurbuchen2016, Lepri2021}, but recent analyses have also made significant progress quantifying the helium variability during ambient solar wind intervals \citep{Kasper2007, Suess2009, WangYM2016b, SanchezDiaz2019}. For example, \citet{Kasper2007, Kasper2012} have shown the solar wind $\alpha/p$ ratio exhibits both a dependence on solar wind speed and the phase of the solar activity cycle, with the $A_{\rm He}$ in the slowest speed solar wind intervals showing the most variation with sunspot number, in support of multiple sources and/or mechanisms for the solar wind's helium component \citep{Schwenn2006}. \citet{Viall2009b} and \citet{Kepko2016} and others have shown that the solar wind helium abundance (and the associated increase in the variance of the helium abundance) are often coincident with periodic proton density structures (and their increased variance), as well as periods of increased ionic and elemental composition \citep[see also][]{Kasper2012}.

Figure~\ref{fig:int} shows a plot of the \emph{Wind}/3DP \citep{Lin1995} and \emph{Wind}/SWE \citep{Ogilvie1995} data at 1~AU for Carrington Rotation 2002 (from 2003 Apr~15 21:35~UT through 2003 May 13 03:24~UT). From top-to-bottom, we plot the bulk radial velocity $V_r$, proton number density $n_p$, alpha number density $n_\alpha$, the $A_{\rm He}$ ratio, and its variance, ${\rm Var}[\, A_{\rm He} \,] \equiv \sigma^2_{\alpha/p}$, calculated over 6-hour bins. The 3DP data are shown in black and the SWE data are shown in red. Based on visual inspection of the Figure~\ref{fig:int} time series, we have identified eight distinct intervals during CR 2002 that can be considered slow-to-moderate speed solar wind ($V_r \lesssim 550$~km~s$^{-1}$) with one or more of the following: enhanced proton density ($n_p \ge 5$~cm$^{-3}$); enhanced alpha density ($n_\alpha \ge 0.25$~cm$^{-3}$); enhanced $A_{\rm He}$ ($\ge 5$\%); or enhanced $\sigma^2_{\alpha/p}$ ($\ge 0.80$). Each of the intervals are labeled above the top $x$-axis as  \#1--8 and shaded as yellow, green, teal, or purple. The colors were selected to represent different large-scale coronal source region configurations, as will be discussed in Section~\ref{sec:map:sweb}. The one exception to our slow-to-moderate speed criteria is interval \#2 (shaded purple) which is clearly identified as a fast ICME, and cataloged as such by \citet{Richardson2010}. The start and end times of each Figure~\ref{fig:int} interval are listed in Table~\ref{tab:int} along with a synopsis of the relevant interval-averaged quantities.

%
\begin{table*}[!bth]
\footnotesize
\centering
 \begin{tabular}{|c|cc|c|rrrrrrr|} 
 \hline
 & Start time & End time & Source & $\langle V_r \rangle$ & $\langle n_p \rangle$ & $\langle A_{\rm He} \rangle$ & $\langle Q_{\rm C} \rangle$ & $\langle Q_{\rm O} \rangle$ & $\langle Q_{\rm Fe} \rangle$ & $\langle {\rm Fe}/{\rm O} \rangle$ \\ 
 \#  & \multicolumn{2}{c|}{DD/MM HH:MM [UT]} & region & [km~s$^{-1}$] & [cm$^{-3}$] & [\%] & [4--6+]& [5--8+] & [6--20+] & \\[1.0ex] 
 \hline
 8  & 04/19 13:58 & 04/21 05:24 & HS (Y)& ${\bf 556 \pm 26}$ & ${\bf 6.1 \pm 4.1}$ & ${\bf 5.1 \pm 1.6}$  & ${\bf 5.07 \pm 0.15}$ & ${\bf 6.14 \pm 0.08}$ & ${\bf 10.82 \pm 0.96}$ & $0.11 \pm 0.02$ \\ 
 7  & 04/23 00:00 & 04/24 14:12 & PS (G)& ${\bf 498 \pm 32}$ & ${\bf 5.2 \pm 1.3}$ & $4.4 \pm 1.6$ & ${\bf 5.19 \pm 0.08}$ & ${\bf 6.08 \pm 0.02}$ & $10.12 \pm 0.52$ & ${\bf 0.13 \pm 0.02}$ \\
 6  & 04/25 14:24 & 04/27 19:12 & PS (G)& ${\bf 478 \pm 34}$ & ${\bf 5.2 \pm 1.1}$ & ${\bf 5.0 \pm 1.1}$ & ${\bf 5.24 \pm 0.16}$ & ${\bf 6.09 \pm 0.05}$ & $9.85 \pm 0.32$ & ${\bf 0.16 \pm 0.03}$ \\
 5  & 04/28 04:48 & 04/29 04:47 & PS (T)& ${\bf 432 \pm 45}$ & $2.2 \pm 1.0$ & $3.9 \pm 1.8$ & ${\bf 5.52 \pm 0.18}$ & ${\bf 6.32 \pm 0.16}$ & ${\bf 11.83 \pm 1.29}$ & ${\bf 0.19 \pm 0.09}$ \\
 4  & 04/29 04:48 & 04/30 09:36 & PS (T)& ${\bf 534 \pm 32}$ & ${\bf 4.5 \pm 2.3}$ & ${\bf 7.6 \pm 2.4}$ & ${\bf 5.35 \pm 0.15}$ & ${\bf 6.29 \pm 0.14}$ & ${\bf 11.07 \pm 0.51}$ & ${\bf 0.31 \pm 0.28}$ \\
 3  & 05/03 15:27 & 05/06 03:56 & HS (Y)& ${\bf 496 \pm 95}$ & ${\bf 7.9 \pm 2.8}$ & ${\bf 4.6 \pm 1.7}$ & ${\bf 5.24 \pm 0.24}$ & ${\bf 6.20 \pm 0.15}$ & ${\bf 11.01 \pm 0.79}$ & ${\bf 0.14 \pm 0.07}$ \\
 2  & 05/09 04:48 & 05/10 16:48 & ICME (P)& $738 \pm 89$ & $3.5 \pm 2.3$ & $3.2 \pm 2.1$ & ${\bf 5.06 \pm 0.15}$ & ${\bf 6.12 \pm 0.05}$ & ${\bf 11.10 \pm 1.14}$ & ${\bf 0.21 \pm 0.11}$ \\
 1  & 05/10 16:48 & 05/11 12:00 & PS (G)& ${\bf 601 \pm 20}$ & $3.6 \pm 1.8$ & ${\bf 9.4 \pm 5.3}$ & ${\bf 5.19 \pm 0.08}$ & ${\bf 6.22 \pm 0.08}$ & $10.09 \pm 0.25$ & ${\bf 0.20 \pm 0.07}$ \\[1ex] 
 \hline
 \multicolumn{4}{|c|}{Non-interval CR 2002 averages} & $637 \pm 93$ & $4.0 \pm 1.9$ & $4.5 \pm 1.2$ & $5.01 \pm 0.17$ & $6.04 \pm 0.09$ & $10.37 \pm 0.55$ & $0.11 \pm 0.03$ \\[1ex] 
 \hline
 \end{tabular}
 \caption{The start and end times of each slow-to-moderate speed, composition-enhanced solar wind intervals during CR~2002 along with the interval-averaged solar wind plasma quantities: $V_r$ and $n_p$ (from \emph{Wind}/3DP), $A_{\rm He}$ (from \emph{Wind}/SWE), and $Q_{\rm C}$, $Q_{\rm O}$, $Q_{\rm Fe}$, and Fe/O (from ACE/SWICS). The interval shading is also indicated (Y---yellow, G---green, T---teal, and P---purple). Boldface values are slower/more enhanced than the non-interval averages over the remainder of CR~2002.}
 \label{tab:int}
\end{table*}

\subsection{Ionic and Elemental Composition Enhancement}
\label{sec:int:ace}

Figure~\ref{fig:comp} shows ACE measurements for the CR~2002 solar wind. From top-to-bottom, we plot the SWEPAM \citep{McComas1998b} measurements of the bulk solar wind speed $V_r$, the normalized 272~eV suprathermal electron pitch angle distribution (PAD), the MAG \citep{Smith1998} measurements of $\boldsymbol{B}$ in RTN coordinates and the magentic field orientation angles ({$\delta$ is the elevation angle above/below the RT plane; $\lambda$ is the azimuth angle within the RT plane}), and the SWICS \citep{Gloeckler1998} composition measurements of select ion charge states of carbon ($Q_{\rm C}$: 4--6+), oxygen ($Q_{\rm O}$: 5--8+), and iron ($Q_{\rm Fe}$: 6--20+), as well as the Fe/O abundance ratio. Here, the solar wind speed and magnetic field values are 1-hr averages whereas the SWICS composition measurements are 2-hr averages.

Figure~\ref{fig:comp} also shows each of the slow-to-moderate speed solar wind intervals associated with enhanced $n_p$, $n_\alpha$, or $A_{\rm He}$ variability that were identified in the \emph{Wind} data of Figure~\ref{fig:int}. With the inclusion of the magnetic field and suprathermal electron PAD, the intervals corresponding to sector boundaries and heliospheric current sheet/plasma sheet (HCS/HPS) crossings are immediately apparent as \#8 and \#3, both shaded light yellow.

Another particularly noteworthy feature of Figure~\ref{fig:comp} is that each of the remaining slow-to-moderate speed intervals are coincident with broader suprathermal electron PADs and/or elevated charge states in C, O, and Fe. While recent analyses by \citet{Borovsky2020,Borovsky2021a} have shown that changes in the suprathermal electron strahl intensities often occur with simultaneous changes in other plasma and/or composition properties, here we note that the broader PADs of intervals \#7, \#6, and \#4 exhibit remarkable, qualitative agreement with the synthetic PAD distribution constructed by \citet{Aslanyan2022} from their MHD simulation of interchange reconnection pseudostreamer outflow (lower panels of Figure~\ref{fig1}(c)). We will show in the next section these intervals do, in fact, map to coronal pseudostreamer source regions.

The charge state and elemental composition enhancements during each of the identified intervals have the following properties. The presence of increased C$^{6+}$ and decreased C$^{5+}$ will result in a substantial increase in the C$^{6+}$/C$^{5+}$ ratio which has similar properties to the O$^{7+}$/O$^{6+}$ ratio commonly used to identify periods of increased coronal electron temperatures \citep[e.g.][]{Landi2012c,Kepko2016}. Additionally, every interval except \#6 and \#7 also show a significant increase in O$^{7+}$ along with a corresponding decrease in O$^{6+}$, providing local maxima of the well-known O$^{7+}$/O$^{6+}$ ratio \citep[e.g.][]{ZhaoL2009,WangYM2016a}. During intervals \#3, \#4--5, and \#8, there are also enhanced levels of the higher iron charge states, Fe$^{\ge 12+}$, including some traditionally ``hot'' signatures of Fe$^{\ge 16+}$ \citep{Lepri2004}. Finally, the elemental composition ratio Fe/O shows clear enhancements during intervals \#1--5 but less obvious enhancement during intervals \#6--7. From Table~\ref{tab:int}, only interval \#8 does not exceed the non-interval Fe/O average.

\section{Solar--Heliospheric Connectivity to the Coronal S-Web}
\label{sec:map}

\begin{figure*}[!t]
	\centerline{ \includegraphics[width=0.90\textwidth]{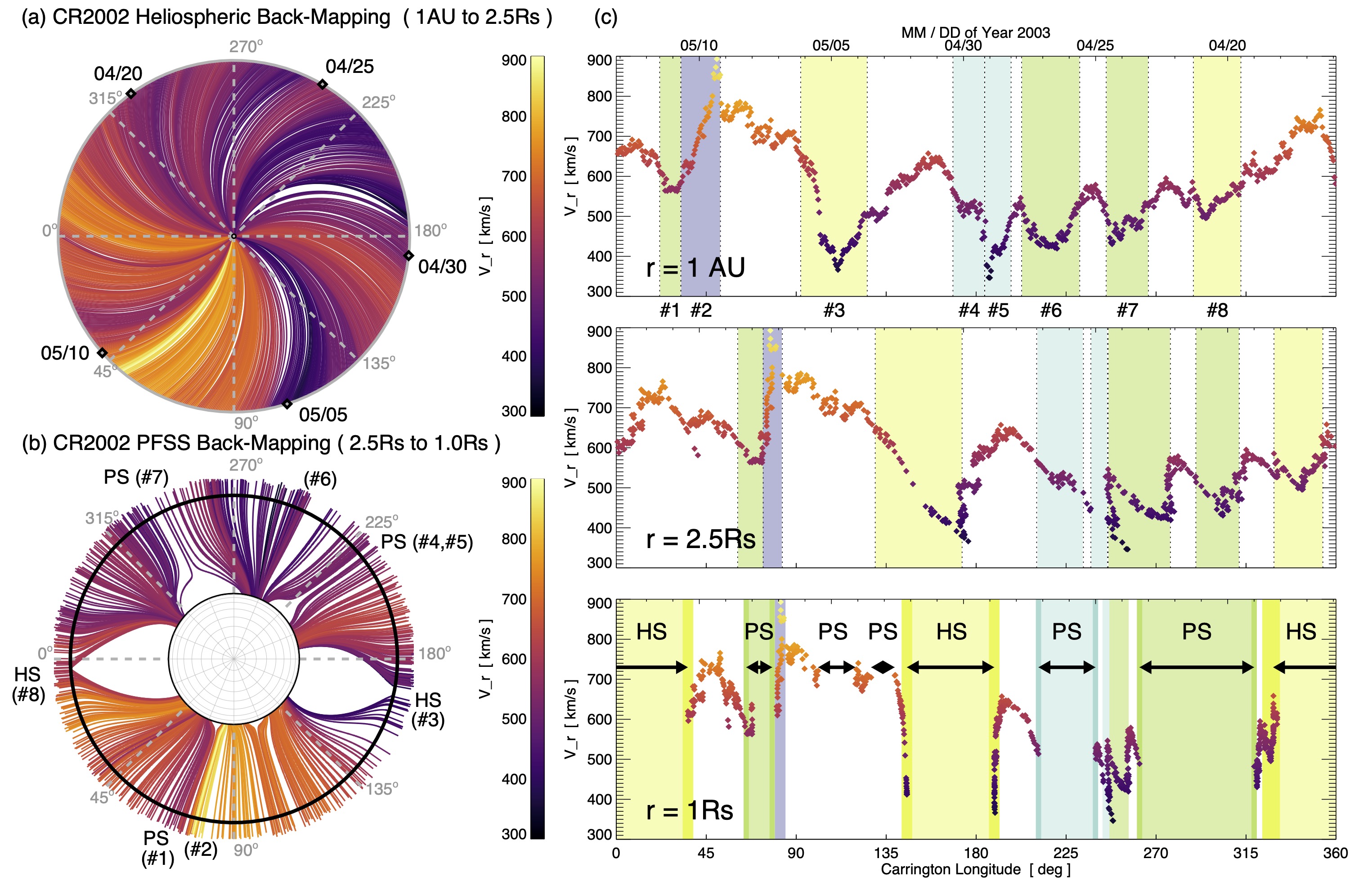} }
	\caption{Heliospheric back-mapping for CR2002. (a) Ecliptic plane streamlines color-coded by 1~AU radial velocity value to the $r=2.5R_\odot$ source surface. (b) Continuation of the back-mapping from $R_{ss}$ to $1R_\odot$ with the PFSS magnetic field. The view of the ecliptic plane is from solar north pole. (c) The mapping of the time series of 1-hr ACE/SWEPAM radial velocity in Carrington Longitude at 1~AU (top panel) to the source surface (middle panel) and then to the solar surface (bottom panel). The intervals of high $\alpha/p$ from Section~\ref{sec:int} are also shown in each location.}
	\label{fig:backmap}
\end{figure*}

\subsection{Heliospheric Ballistic Back-Mapping}
\label{sec:map:heliosphere}

Here we follow the standard procedure for heliospheric backmapping described by \citet{Parenti2021} and references therein. The in-situ observations of solar wind at 1~AU are ballistically mapped from the spacecraft back to the Sun along the \citet{Parker1958} spiral streamlines assuming constant $V_r$ values equal to the 1~hr averages measured by ACE. Figure~\ref{fig:backmap}(a) plots the heliospheric representation of Parker spiral streamlines colored by radial velocity. 

In order to map the in-situ solar wind observations back to their coronal source regions on the solar surface, we use the standard PFSS model \citep[e.g.][]{Altschuler1969,WangYM1992} to approximate the large-scale geometry of the solar corona. We calculate the PFSS extrapolation from the line-of-sight observations of the photospheric magnetic field taken by MDI \citep{Scherrer1995} aboard the \emph{Solar and Heliospheric Observatory} \citep[SOHO;][]{Domingo1995}. The line-of-sight fields are transformed into radial fields via the $B_r = B_{\rm los}/\sin{\theta}$ relation. Starting with the original high-resolution ($3600 \times 1080$) MDI synoptic map for CR 2002 with the \citet{Sun2011} interpolation for the polar field values, we rebin it to $720\times360$ and calculate the PFSS spherical harmonics through order $l_{\rm max} = 16$ with a source surface height of $R_{ss}=2.5\,R_{\odot}$. Figure~\ref{fig:backmap}(b) plots the magnetic field line mapping from $R_{ss}$ to the lower boundary at $r=1\,R_{\odot}$. Here the large-scale, closed-field coronal HS and PS structures are labeled with their corresponding intervals.

The top panel of Figure~\ref{fig:backmap}(c) plots the radial velocity as a function of Carrington longitude at 1~AU (note time now runs from right-to-left as indicated by the upper $x$-axis label). We have also drawn the corresponding intervals of enhanced variability identified in $\S$\ref{sec:int:wind}. The middle panel of Figure~\ref{fig:backmap}(c) plots the 1~hr ACE velocity measurements as a function of Carrington longitude at $R_{ss}$ while the bottom panel of Figure~\ref{fig:backmap}(c) shows the Carrington longitude of the streamline footpoints at $1\,R_{\odot}$.

This ballistic mapping method has been widely used to estimate the coronal source regions of in-situ solar wind measurements \citep[e.g.][]{Neugebauer2002, Neugebauer2004, Gibson2011, ZhaoL2013a, ZhaoL2017a}, including with the new PSP \citep[e.g][]{Badman2020,Panasenco2020,Griton2021} and \emph{Solar Orbiter} data \citep[e.g.][]{Telloni2021}. We note that, while the numerical errors associated with integrating velocity streamlines or magnetic field lines, e.g. from a PFSS extrapolation, are quite small \citep{Stansby2022}, the overall ``uncertainty'' in the position of the foot point of the magnetic field lines as mapped by these techniques are typically within approximately 10$^{\circ}$ \citep{Neugebauer2002, Leamon2009}, largely due to the assumptions and simplifications inherent in the models themselves, such as the current-free approximation in the corona and the unperturbed Parker spiral structure that does not account for the interaction between fast and slow solar wind streams, etc.

\subsection{S-Web Source Region Configurations}
\label{sec:map:sweb}

The static representation of the Separatrix Web (S-Web) topological structures is based on the $Q$-map which is defined as the logarithmic ``squashing factor,'' $\log{Q}$. The $Q$-map quantifies the magnetic field's geometric connectivity \citep{Titov2007}, i.e., separatrix and quasi-separatrix surfaces are regions of high $Q$  \citep[e.g.][]{Titov2011,Antiochos2012,Scott2018}. We have calculated the $Q$ value from the CR~2002 PFSS magnetic field extrapolation via the formulation in \citet{Titov2007} where $Q = N^2 / |\Delta|$,  

\begin{equation}
N^2 =  
\left(\frac{\partial Y}{\partial y}\right)^2 +
\left(\frac{\partial Y}{\partial z}\right)^2 + 
\left(\frac{\partial Z}{\partial y}\right)^2 +
\left(\frac{\partial Z}{\partial z}\right)^2 \; ,
\end{equation}
and $|\Delta| = |B_x/B_x^{*}|$. While the full derivation (in arbitrary coordinates) is described in \citet{Titov2007}, the expression in spherical coordinates is straightforward to obtain: $B_x/B_x^{*} \rightarrow B_r/B_r^{*}$, the starting and ending field line positions become $(x_0, y_0, z_0) \rightarrow (r_0, \theta_0, \phi_0)$, $(X,Y,Z) \rightarrow (R,\Theta,\Phi)$, and the differentials become changes in arc length  
$\partial y \rightarrow r_0 \, \partial \theta$, 
$\partial Y \rightarrow R \, \partial \Theta$, 
$\partial z \rightarrow r_0  \sin{\theta_0} \, \partial \phi$, and 
$\partial Z \rightarrow R  \sin{\Theta} \, \partial \Phi $. 
We note that when the starting and ending radial surfaces are set to $r_0 = R_\odot$ and $R=R_{*}$, one arrives at the exact spherical definition of $N^2$ given as Equation (22) in \citet{Titov2007}. We calculate the field connectivity from a grid of $1536 \times 768$ field lines starting at the desired radial distance $r_0$ uniformly spaced in ($\theta$, $\phi$). As in the $Q$-map calculation of \citet{Wyper2016b}, we use a fourth-order central difference scheme for the derivatives.

\begin{figure}[!t]
    \includegraphics[width=0.47\textwidth]{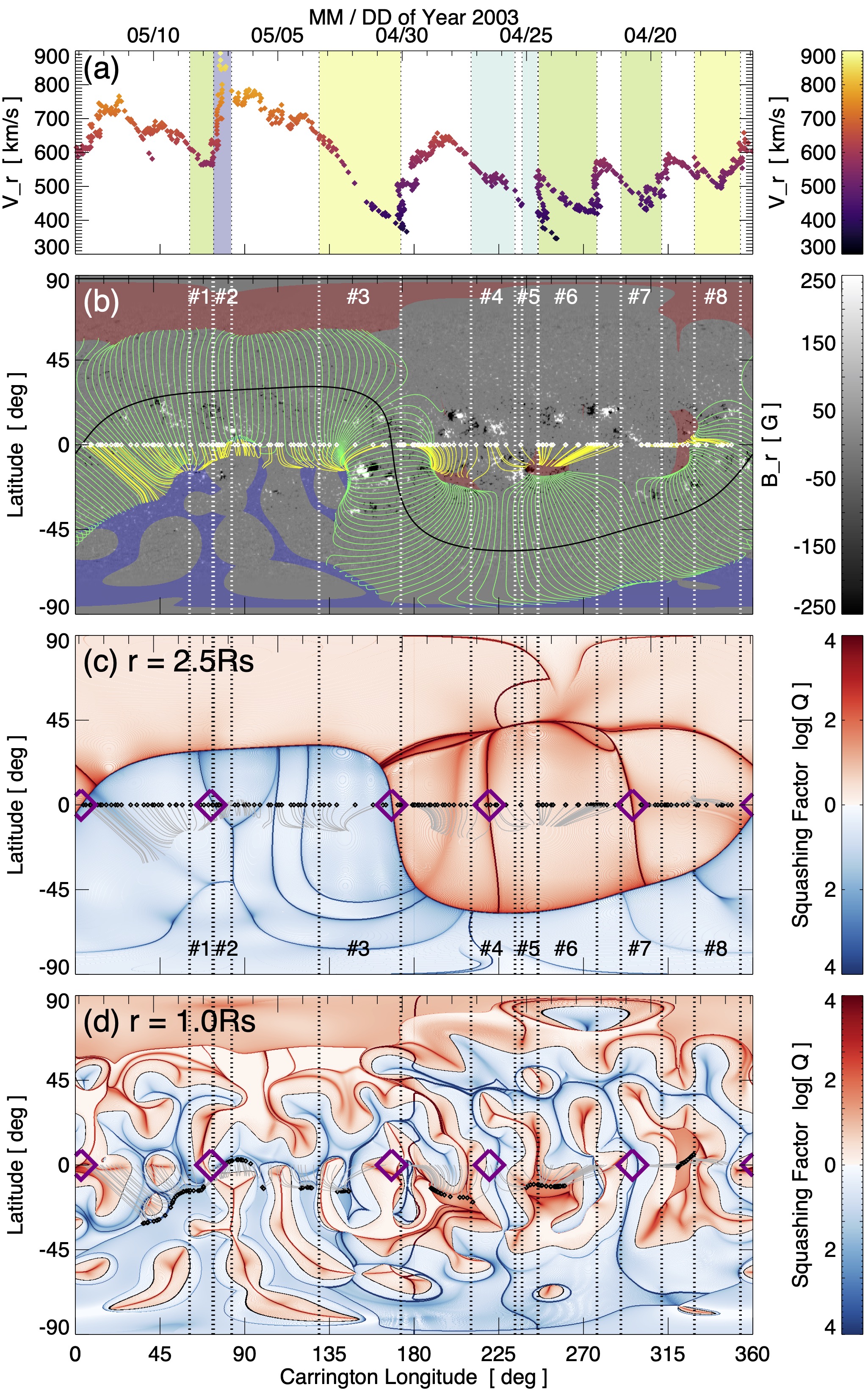}
	\caption{Magnetic structure of the PFSS extrapolation for CR2002 with back-mapped in-situ intervals of slow-to-moderate speed solar wind. (a) The back-mapped $V_r$ time series and intervals \#1--8 at $R_{ss}=2.5R_{\odot}$ from Figure~\ref{fig:backmap}(c). (b) Synoptic map of the open field regions (blue positive polarity, red negative polarity). The configuration of the helmet streamer belt is shown as green field lines traced from the $B_r=0$ contour at $R_{ss}$ representing the location of the HCS. (c) $Q$-map at $R_{ss}$ showing the characteristic arcs of the S-Web structure. The $\log Q$ values are shaded blue (red) for positive (negative) polarity. (d) $Q$-map at $1R_{\odot}$ showing the equatorial field line foot point locations and the low-latitude open field regions between PS's for intervals \#4--7.}
	\label{fig:sweb}
\end{figure}

Figure~\ref{fig:sweb} summarizes the coronal portion of our heliospheric back-mapping procedure to illustrate the connectivity of our composition-enhanced intervals to their coronal S-Web structures of origin. Figure~\ref{fig:sweb}(a) repeats the 1-hr average $V_r$ points mapped to $R_{ss}$ (from Figure~\ref{fig:backmap}(c)) and plots the longitudinal extent of our back-mapped intervals of interest with their boundaries indicated in every subsequent panel.  Figure~\ref{fig:sweb}(b) shows the MDI magnetogram for CR~2002. The positive (negative) open field regions calculated from PFSS solution shaded in blue (red), the structure of the helmet streamer belt with representative green field lines, and the HCS location ($B_r=0$ at $R_{ss}$) as the black contour. The green field lines are traced along the HCS location at a radial distance just below $R_{ss}$, and therefore represent the largest closed flux tubes belonging to the helmet streamer belt and illustrate the boundary between the large-scale open and closed coronal flux systems. The back-mapped intervals are labeled along the top axis of the plot.

Figure~\ref{fig:sweb}(c) plots the $Q$-map at $R_{ss}$. The values of $\log{Q}$ are also shaded blue and red to indicate $B_r$ polarity. The position of the HCS current sheet is immediately identified as where the polarities change sign. The darker arcs contained within each polarity correspond to S-Web arcs. These S-Web arcs indicate the PFSS field line mapping of the outer spine and fan structures of PS flux systems \citep{Scott2018} and/or the presence of narrow channels of open field \citep{Antiochos2007}. The purple diamonds indicate the S-Web features associated with their corresponding back-mapped, composition-enhanced intervals. The in-situ intervals that contain the HCS crossing (\#3, \#8) are clearly associated with the HS belt and the intersection of the HCS with the ecliptic plane, despite the spatial extent of interval \#8 at $R_{ss}$ (329$^{\circ}$--353$^{\circ}$) missing the PFSS location of the HCS (3.7$^{\circ}$) by $\sim$10$^{\circ}$. This discrepancy is typical of the uncertainties associated with our simplified back-mapping (as mentioned above) but given that PFSS helmet streamer width beneath the HCS--ecliptic plane intersection spans $\sim$67$^{\circ}$ in Carrington longitude (330$^{\circ}$--37$^{\circ}$) at $R_{\odot}$, the association between the solar wind during interval \#8 and its origin from this portion of the helmet streamer is evident. Intervals \#1, \#4, and \#7 each include a well-defined, PS S-Web arc in their longitudinal range. Interval \#5 is directly adjacent to the S-Web arc of interval \#4 and interval \#6 appears to straddle the midpoint between the \#4 and \#7 S-Web arcs.     

Figure~\ref{fig:sweb}(d) plots the $Q$-map at $r=1\,R_{\odot}$ and shows the foot points of the PFSS magnetic field lines traced from $R_{ss}$. The positive polarity (blue) open field foot points map to the southern boundary of the HS belt/northern boundary of the polar coronal hole extensions. The negative polarity (red) open field foot points map to a series of low-latitude coronal holes sandwiched between the northern boundary of the HS belt and the southern boundaries of a series of large PSs above the AR magnetic fields between Carrington longitudes 180$^{\circ}$--315$^{\circ}$. While intervals \#5 and \#6 were not associated with a distinct S-Web arc at $R_{ss}$, their field line foot points map to the vicinity of the open--closed flux boundaries between the low-latitude coronal holes and the large equatorial PSs.

\section{Coherent Magnetic Structure in Composition-Enhanced Solar Wind}
\label{sec:results}

\subsection{$H_m$--PVI Analysis Procedure}
\label{sec:results:hmpvi}

We have implemented the \citet{Pecora2021} magnetic helicity--partial variance of increments ($H_m$--PVI) procedure to identify coherent magnetic structures within our intervals of composition-enhanced solar wind originating from coronal HS and PS source regions. Here, we briefly review the methodology for the identification of small-scale flux ropes and/or coherent flux tubes, while in sections~\ref{sec:results:hswind} and \ref{sec:results:pswind}, we present the results from applying this technique to HS intervals (\#3, \#8) and PS intervals (\#1, \#4--\#7), respectively. In section~\ref{sec:results:stats}, we compare and contrast properties of the $H_m$ and PVI time series in each interval.

Quite generally, the magnetic helicity can be written as

\begin{equation}
H_m = H_m^{+}(\ell) + H_m^{-}(\ell)    
\end{equation}
where the temporal or spatial scale, $\ell$, is used to define the magnetic helicity contained in scales greater than $\ell$ as $H_m^{+}(\ell)$ and less than $\ell$ as $H_m^{-}(\ell)$. Since we are interested in the local coherence, we will ignore the $H_m^{+}$ term and follow the \citet{Pecora2021} prescription for the local estimate of $H_m^{-}$ using the two-point correlation function $C_{jk} = \langle \, B_j(\boldsymbol{r}) B_k(\boldsymbol{r}+\boldsymbol{s}) \, \rangle$ \citep[e.g.][]{Matthaeus1982}. We take the spatial lag $\boldsymbol{s} = s \, \hat{\boldsymbol{e}}_i$ to be in the $\hat{\boldsymbol{r}}$ direction so indices $j$, $k$ are the tangent and normal directions of the spacecraft's RTN coordinates.

We calculate the spatial average of the two-point correlation function over an interval of width $W = 2 \ell$ centered at position $x$ via

\begin{equation}
\begin{split}
    & \;\;\;\; C_{jk}(x,s) = \\ 
    & \frac{1}{W} \int_{x-\frac{W}{2}}^{x+\frac{W}{2}} d\xi \, \Big[ B_{j}(\xi + s) B_{k}(\xi)  - B_{j}(\xi) B_{k}(\xi+s) \Big] \, .
\end{split}
\end{equation}
Following \citet{Pecora2021}, we apply a smooth windowing function to $C_{jk}(x,s)$ of the form $f(s) = \onehalf + \onehalf \cos{\left( \, 2 \pi s / W \, \right)}$ to obtain the local helicity estimate,

\begin{equation}
    H_m(x,\ell) = \int_0^\ell ds \; C_{jk}(x,s) \; f(s) \; .
\end{equation}
The spatial domain quantities ($x$, $s$) can be converted to the temporal domain ($t$, $\tau$) with the usual Taylor approximation of $x(t) = \int d\tau \; V_r(\tau)$.

In our implementation of $H_m(t,\ell)$, we use a spatial scale of $\ell_H = 4.3 \times 10^{6}$~km ($4300$~Mm, ${\sim}6.2\,R_{\odot}$) and for a solar wind speed of $500$~km~s$^{-1}$, this corresponds to a temporal scale of $2.4$~hr (i.e. ${\sim}135$~data points at $64$~s cadence) which is the mean correlation timescale of the vector magnetic field over our eight intervals ($2.37 \pm 1.83$~hrs). However, we note that the correlation timescales during the HCS/HPS intervals were significantly larger ($4.58 \pm 1.0$~hrs) than the PS intervals ($0.98 \pm 0.52$~hrs) which agree with previous estimates \citep[e.g.][]{Matthaeus2005}. Typically, one decides that a given peak in $H_m(t,\ell)$ is significant if it exceeds a $\pm1$-$\sigma$ threshold. In the following sections, this standard deviation is calculated from the $H_m$ curves over the entire interval of interest, i.e. those defined in Section~\ref{sec:int} (and illustrated in Figures~\ref{fig:int}--\ref{fig:sweb}).

The PVI measure \citep[e.g.][]{Pecora2019,Pecora2021} is defined as 

\begin{equation}
    {\rm PVI}(t,\ell) = \frac{|\Delta \boldsymbol{B}(t,\ell)|}{\sqrt{\langle \, |\Delta \boldsymbol{B}(t,\ell)|^2 \, \rangle}} \; ,
\end{equation}

\noindent in which $|\Delta \boldsymbol{B}(t,\ell)| \equiv |\boldsymbol{B}(t+\ell) - \boldsymbol{B}(t)|$, the (temporal or spatial) averaging is over an appropriate interval, and $\ell$ represents the scale size of the increments. The PVI technique has been widely used to identify discontinuities, reconnecting current sheets, and as a measure of turbulence structures \citep[e.g.][]{Greco2009a,Greco2009b,Greco2018,Osman2014,Pecora2019}. Since we aim to use PVI to identify the sharp magnetic boundaries of coherent flux tubes and/or small flux rope plasmoids, we choose a temporal scale of $\ell_{\rm PVI} = 2.13$~min and an averaging window of 24~hrs (10 times the magnetic field's mean correlation timescale above). Again, one makes a determination of the significance of any given PVI peak via thresholding, where some authors have used PVI~$>2$ \citep{Pecora2021}, ${>}\,2.4$ \citep{Greco2008}, ${>}\,3$ \citep{Kilpua2022}, or even larger thresholds of ${>}\,4$--$6$ \citep[e.g.][]{Servidio2011,Zhou2019}. Here we use PVI~$>3.0$ during each of our CR~2002 solar wind intervals for ease of comparison between the HS and PS PVI statistics. The magnitude of the PVI peaks has been shown to be related to different types of boundaries or discontinuities in the solar wind. For example, the PVI $\gtrsim 3$ threshold has been interpreted as representing discontinuities that are actual physical boundaries of coherent magnetic structures rather than random statistical fluctuations, whereas PVI values $\gtrsim 5$ have been associated with reconnection events \citep{Servidio2011}.

The strength of the $H_m$--PVI procedure is that for a magnetic island or a coherent flux rope-like structure there is a local $H_m(t)$ maximum somewhere within the flux rope and ${\rm PVI}(t)$ yields local maxima at the flux rope boundaries. For a given time series, local peaks in $H_m$ or PVI can each occur for a variety of independent  features, but the combination of two PVI peaks bracketing a local $H_m$ maximum appears to be a fairly robust identification criteria \citep{Pecora2021}.  

\begin{figure*}[!t]
	\centerline{ \includegraphics[width=1.0\textwidth,height=4.35in]{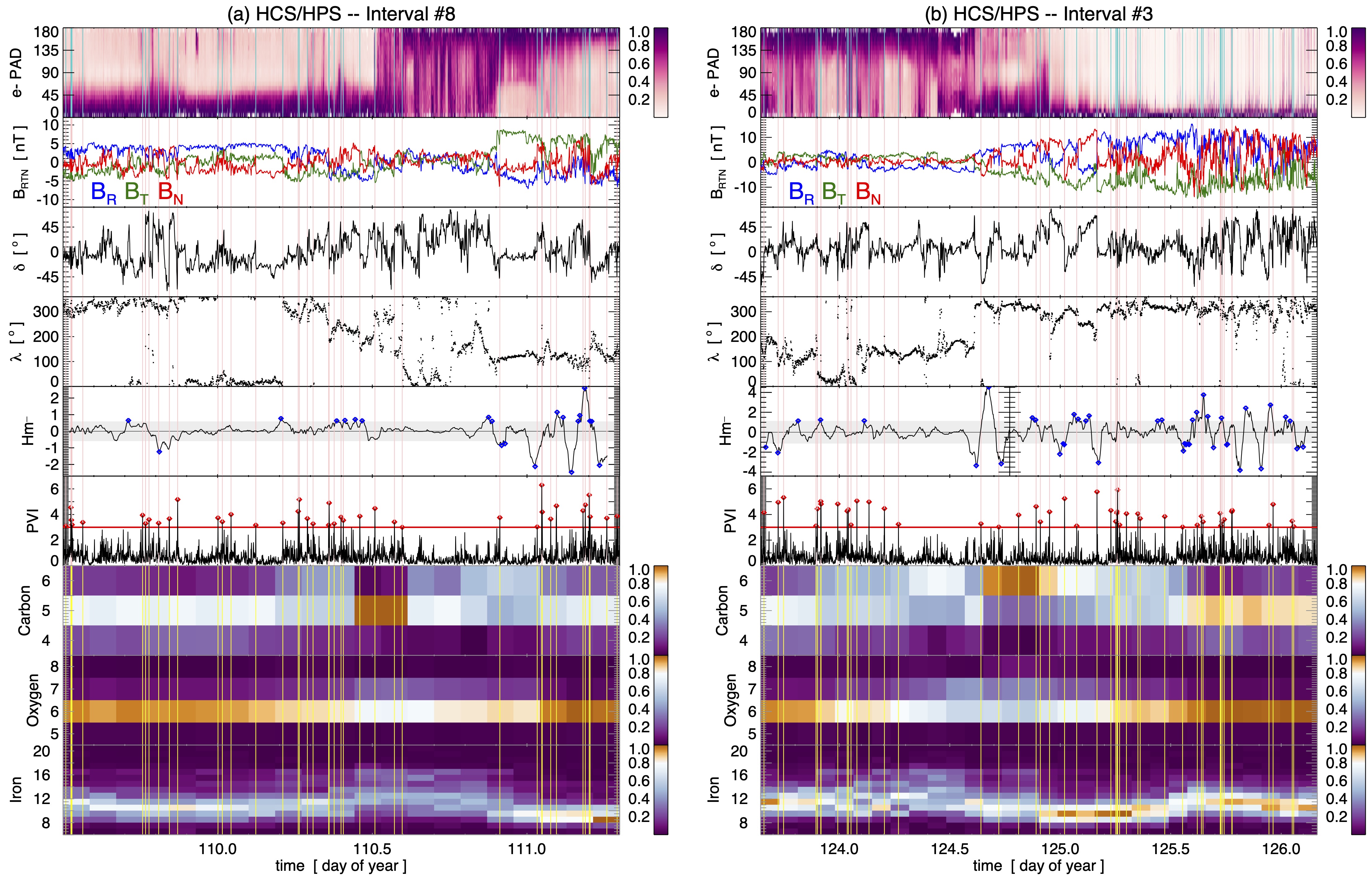} }
	\caption{Intervals of the helmet streamer (HS) wind that include heliospheric current sheet/plasma sheet crossings. (a) HS (\#8) from DOY 109.582 to 111.267. (b) HS (\#3) from DOY 123.644 to 126.164.}
	\label{fig:hcsints}
\end{figure*}

For completeness, we note there are a number of complementary methods to identify coherent intervals and/or solar wind flux tube boundaries based on either statistical plasma properties or turbulence measures. For example, rapid changes in the magnetic field orientation (i.e. tangential discontinuities) can be characterized with $\Delta \theta_{B}$ \citep[e.g.][and references therein]{Borovsky2008}, and these have recently been shown to coincide with abrupt changes in the suprathermal electron strahl width and/or intensity \citep{Borovsky2020, Borovsky2021a, Borovsky2021b}.

\subsection{Intervals of Helmet Streamer (HS) Wind}
\label{sec:results:hswind}

Figure~\ref{fig:hcsints} shows our two composition-enhanced intervals associated with HS wind and the in-situ HCS/HPS and IMF sector boundary crossings. Figure~\ref{fig:hcsints}(a) shows interval \#8 which is from DOY 109.582 to 111.267 (40.44~hr total duration) and Figure~\ref{fig:hcsints}(b) shows interval \#3, from DOY 123.644 to 126.164 (60.48~hr total duration). In each plot, the top panel shows the (normalized) 272~eV suprathermal electron PAD. The next three panels show the 64~s vector magnetic field in RTN components ($B_R$ blue; $B_T$ green; $B_N$ red) along with its local orientation angles: latitude $\delta \in [-90^{\circ}, \, 90^{\circ}]$ and longitude $\lambda \in [0^{\circ}, \, 360^{\circ}$]. The fourth and fifth panels show the magnetic helicity measure $H_m^{-}$ and the PVI profiles. And the remaining three panels show the 2-hr ionic composition measurements from ACE/SWICS of C$^{4-6+}$, O$^{5-8+}$, and Fe$^{6-20+}$.

The PVI panels have the ${\rm PVI} \ge 3.0$ threshold shown as a solid red line and the local maxima are indicated with red diamond symbols. The vertical lines associated with the location of the PVI peaks are drawn over each panel. The $H_m^{-}$ profile in \ref{fig:hcsints}(a) is normalized by a value of $1.0 \times 10^{8}$, and has a standard deviation of $\sigma = 0.582$. The $H_m^{-}$ profile in \ref{fig:hcsints}(b) has two separate normalizations, as indicated by the additional $y$-axis at DOY 124.764 due to the magnitude of $\boldsymbol{B}$ increasing after the HCS/HPS crossing. For $t < 124.764$ the normalization value is $1.0 \times 10^8$ whereas for $t > 124.764$, we normalize by $7.68 \times 10^8$ so that $\sigma = 1.124$ for both sides. In each of the $H_m^{-}$ panels the $\pm 1 \sigma$ range is shaded light gray. The local $H_m^{-}$ magnitude maxima larger than 1-$\sigma$ within each PVI interval are indicated with blue diamond plot symbols.  

The $H_m$--PVI procedure identifies a number of coherent magnetic structures throughout each interval. The occurrence of significant PVI peaks tend to be clustered in local patches and each interval's HCS/HPS crossing (the $\sim$180$^\circ$ transition in $\lambda$ coincident with the bidirectional/broadening of the suprathermal PADs) is bracketed by a cluster of PVI-peaks.

In HS interval \#8, there are four main clusters of PVI peaks: DOY 109.7--109.9, 110.0--110.15, 110.2--110.6, and 110.9--111.3. The two largest clusters of PVI peaks occur on either side of the HCS/HPS crossing and contain the greatest number of significant $H_m$ peaks. Once the suprathermal electron PAD transitions from a unidirectional ($0^{\circ}$) strahl to a broader, more isotropic distribution (DOY 110.5 through 111.1) there is a train of three coherent north-to-south magnetic field rotations (positive-to-negative profile in $\delta$) at the beginning of the PAD transition and a number of larger $H_m$ structures as the PAD transitions to oppositely-directed ($180^{\circ}$) strahl on the other side. Notably, the structure centered at DOY 111.0 corresponds to a 1.5~hr-wide, relatively flat profile in both $\delta$ and $\lambda$. Finally, there is a slight increase in O$^{7+}$ (and decrease in O$^{6+}$) for the duration of the HCS/HPS crossing during the broad electron PAD region which coincides with a slight shift to higher Fe charge states during this same period. Likewise, there is a significant increase in C$^{6+}$ at the beginning and end of the PAD transitions at the same time as a number of the HCS/HPS interval-related $H_m$--PVI structures.

HS interval \#3 exhibits many similar features to those of interval \#8. For example, the significant $H_m^{-}$ peaks occur on either side of the IMF sector boundary at DOY 124.6, including three consecutive structures between DOY 124.5--124.8, followed by three more, centered on DOY 125.0, 125.1, and 125.2. These $H_m$--PVI structures are also associated with coherent rotations in $(\delta, \lambda)$, as well as a sharp local maximum in C$^{6+}$ at the HCS superimposed on top of a broader region of enhanced C$^{6+}$ from DOY 124.0--125.5. The O$^{7+}$ signal shows a similar, but less pronounced, trend over a slightly narrower range (DOY 124.3--125.2). However, the enhanced high Fe charge states (up to Fe$^{16+}$) tend to be shifted earlier (DOY 123.9--124.6) and return to being strongly peaked at Fe$^{9-11+}$ for $t > 125.0$. The suprathermal electron PADs leading up to the HCS crossing are more patchy, alternating between the 180$^{\circ}$ strahl and broader, more isotropic (and even bidirectional) PADs before transitioning to more steady 0$^{\circ}$ strahl after DOY 125.0.

Throughout both intervals, the PVI peaks occur almost exclusively at discontinuities in the magnetic field angles, and these are often also coincident with changes in the electron PADs. Thus, the conjecture that the PVI peaks select boundaries of distinct plasma intervals (either magnetic flux tubes, discrete solar wind flows, or magnetic island plasmoid/small flux ropes) appears supported by our results. Another feature of the $H_m$--PVI analysis in these intervals is that, even when the $H_m$ profiles do not exceed the 1-$\sigma$ significance threshold, there are often still local peaks and coherent magnetic field signatures within the bracketing PVI peaks.

\subsection{Intervals of Pseudostreamer (PS) Wind}
\label{sec:results:pswind}

\begin{figure*}[!t]
	\centerline{\includegraphics[width=1.0\textwidth,height=4.35in]{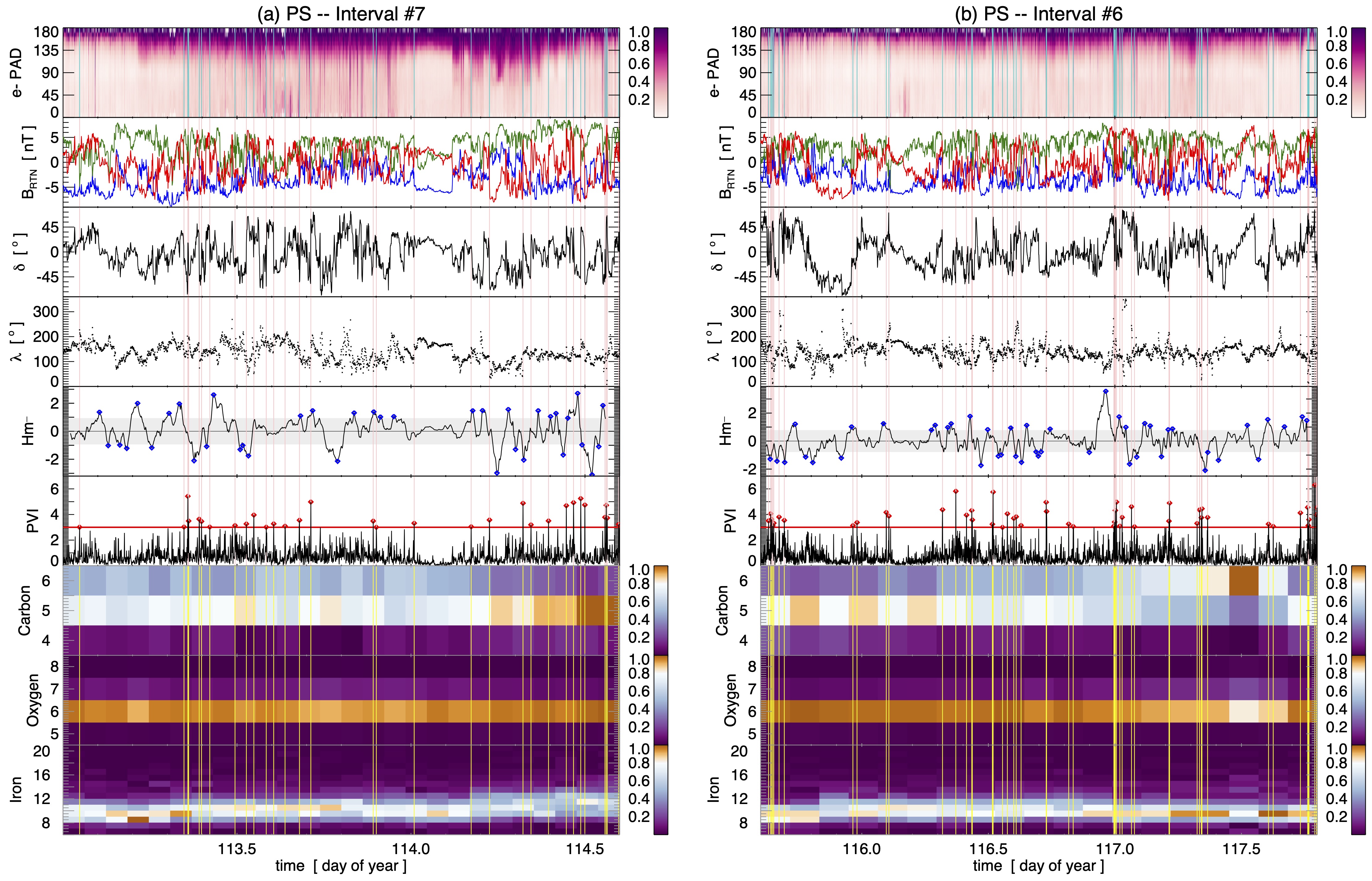}}
	\centerline{\includegraphics[width=1.0\textwidth,height=4.35in]{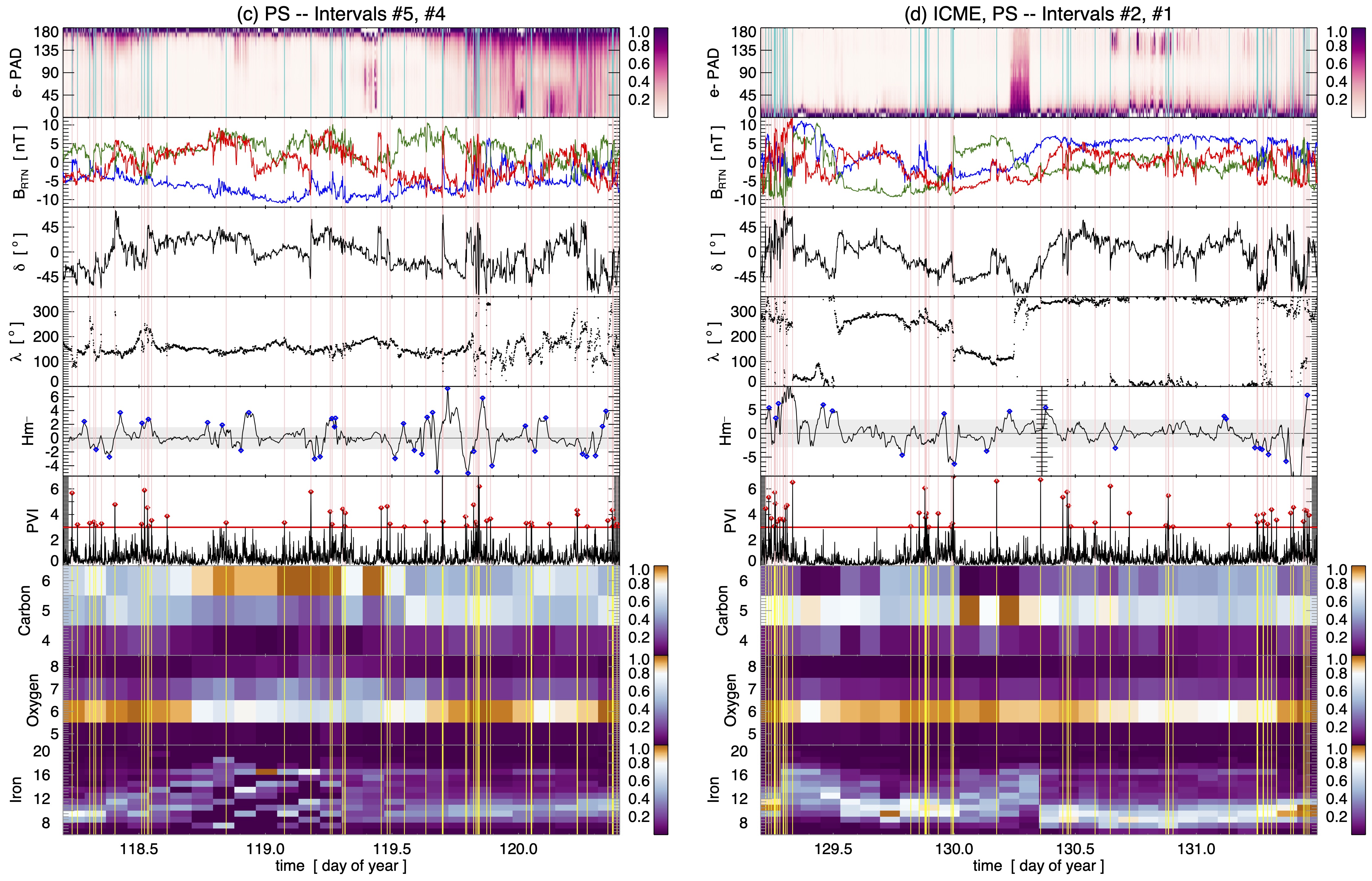}}
	\caption{Non-helmet streamer, composition-enhanced intervals (primarily from pseudostreamers) in the same format as Figure~\ref{fig:hcsints}. (a) PS (\#7) from DOY 113.0 to 114.60. (b) PS (\#6) from DOY 115.60 to 117.80. (c) PS (\#5, \#4) from DOY 118.20 to 120.40. (d) ICME (\#2) and PS (\#1) from DOY 130.70 to 131.50.}
	\label{fig:psints}
\end{figure*}

Figure~\ref{fig:psints} shows the remaining composition-enhanced intervals associated with non-HS wind, i.e. from PS or PS-adjacent source regions, in the same format as Figure~\ref{fig:hcsints}. Figure~\ref{fig:psints}(a) shows interval \#7 which is from DOY 113.0 to 114.6 (38.4~hr duration), Figure~\ref{fig:psints}(b) shows interval \#6, from DOY 115.6 to 117.8 (52.8~hr), Figure~\ref{fig:psints}(c) shows the combined intervals of \#4 and \#5 from DOY 118.2 to 120.4 (52.8~hr), and lastly Figure~\ref{fig:psints}(d) shows the ICME interval \#2 from DOY 129.2 to 130.7 (36~hr) and the subsequent, brief PS interval \#1, from DOY 130.7 to 131.5 (19.2~hr). 

The qualitative features of the HS intervals described above are also present in each of the PS intervals. Specifically, the $H_m$--PVI analysis continues to identify magnetic field discontinuities and/or sudden changes in the electron PAD via the significant PVI peaks, the PVI peaks clearly show clustering, and these peaks often bracket significant local maxima in the $H_m^{-}$ magnitude. The $H_m$ normalization for intervals \#7, \#6 is $1.0 \times 10^8$ which results in a standard deviation of $\sigma = 0.940$ for \#7 and $\sigma = 0.768$ for \#6.  

In interval \#7 (Figure~\ref{fig:psints}(a)), there are a series of significant $H_m$ peaks from DOY 113.1--113.55 that begin before the cluster of PVI $>3$ events ranging from DOY 113.35--113.75 and another succession of $H_m$ peaks coincident with the next large PVI cluster at $t \gtrsim 114.2$. During interval \#6 (Figure~\ref{fig:psints}(b)), the PVI clusters are more frequent and of shorter duration, whereas the significant $H_m$ peaks are more spread out over the entire interval. The overlap between the two occur primarily for DOY 116.3--116.7 and for $t > 116.9$. Essentially the entire \#7 interval has a moderate enhancement of C$^{6+}$, but almost no corresponding enhancement in the hotter charge states of O or Fe. Interval \#6 is similar with perhaps a very slight enhancement in Fe$^{10-12+}$ between DOY 116--117, and a more prominent C$^{6+}$ region for $t > 117$. Additionally, there is one 2-hr data point centered at 117.5 that include a slight increase in O$^{7+}$ (and decrease in O$^{6+}$), coincident with a coherent magnetic structure interval. In general, PS intervals \#7 and \#6 can be considered to have fewer composition enhancement than our previous HS intervals. And while there are still some discrete regions of broader electron PAD signatures, especially in \#7, for the most part these PS intervals have less variation in the PAD profiles---as may be expected for unipolar PS solar wind. 

PS intervals \#5+4 and \#2+1, shown in Figures~\ref{fig:psints}(c) and \ref{fig:psints}(d), respectively, also show PVI clusters and sequential trains of $H_m$ coherent structures bracketed by PVI peaks. The $H_m$ normalization in interval \#5+4 is $1.0 \times 10^8$, resulting in a standard deviation of $\sigma = 1.608$. We use the same normalization ($10^8$) for the ICME interval (\#2) which yields $\sigma = 2.902$ while for trailing PS interval (\#1), we use a normalization of $3.877 \times 10^{7}$ to obtain the matching $\sigma$ value.

PS interval \#5+4 has the most enhanced heavy ion charge states of our PS intervals, including a significant increase in C$^{6+}$ and O$^{7+}$ from DOY 118.7--119.6. coincident with $H_m$ peaks at DOY 118.9,  119.1, 119.2, and 119.3. Interval \#5+4 also contains highly variable and enhanced hot iron charge states, Fe$^{\ge12+}$, including a Fe$^{16+}$ component present throughout almost the entire time range, DOY 118.5--120.2. Additionally, there are (small) flux rope-like rotations in the DOY 118.4--118.5 and 119.7--119.8 structures. Again, the PVI peaks representing coherent structure boundaries are seen to line up with discontinuities in magnetic field $(\delta, \lambda)$ angles. 

While there is interesting, composition-enhanced internal magnetic structuring present within the ICME interval of Figure~\ref{fig:psints}(d)---including large ICME boundary enhancements in the Fe distribution (e.g. DOY 129.3--129.7 and 130.0--130.4)---in this work we will concentrate on the PS interval \#1. The Fe$^{16+}$ component is also present for a large percentage of this interval, through DOY 131.3. There is an intriguing sequence of short, intermittent bursts of bidirectional electrons from DOY 130.6--131.0 which have corresponding $H_m$ structures that do not exceed the $1$-$\sigma$ threshold but occur toward the latter portion of an extended PVI peak cluster. The $H_m$ peaks that do exceed the threshold occur towards the end of interval \#1 and into the beginning of interval \#2 (DOY 130.2--130.7) and the coherent magnetic structures at $t \gtrsim 131.3$ also show flux rope-like rotation signatures.

\subsection{Statistical Properties}
\label{sec:results:stats}

\begin{figure*}[tbh]
	\centerline{\includegraphics[width=1.0\textwidth]{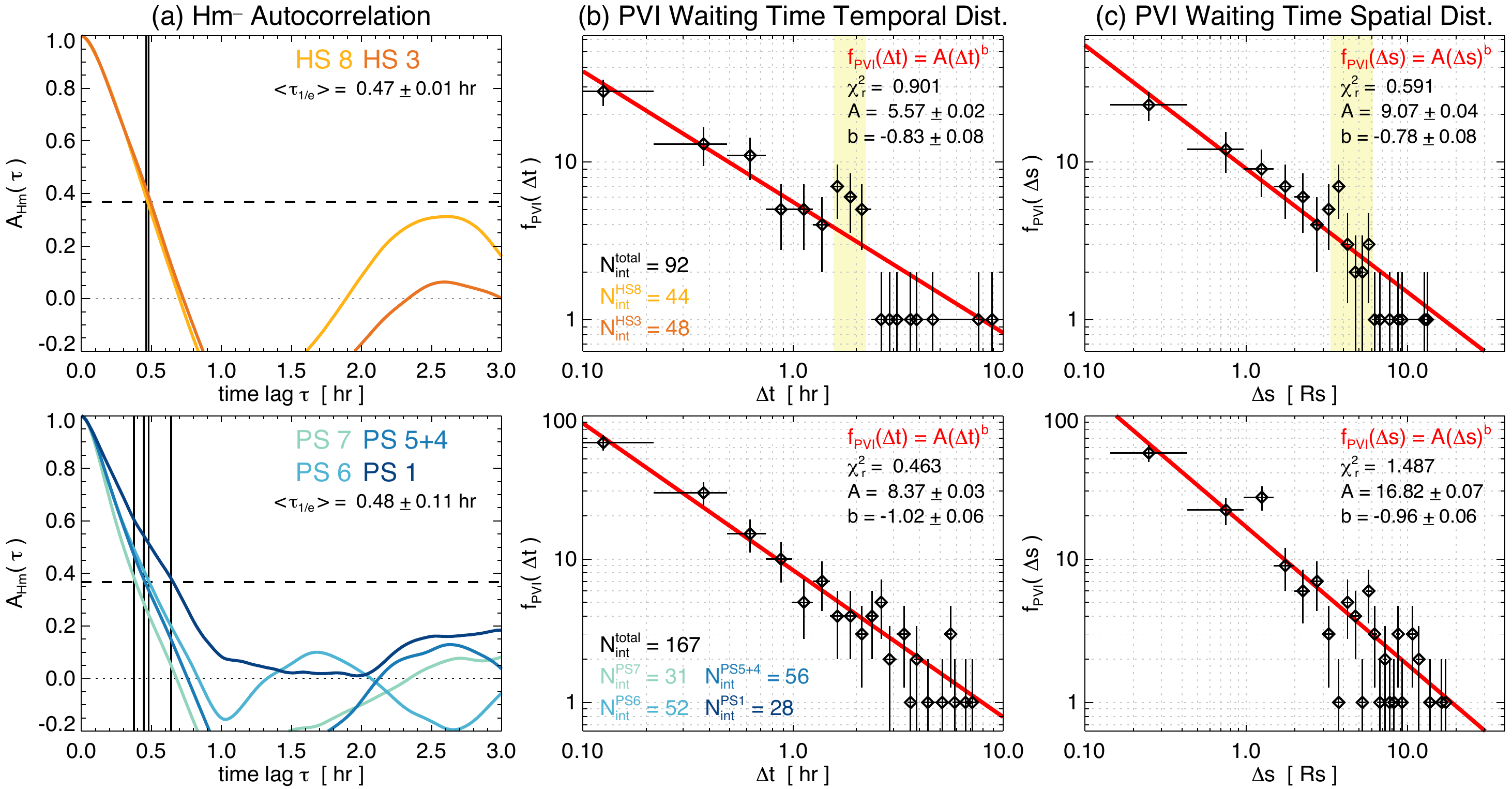}}
	\caption{Statistical properties of the intermittency and coherent magnetic structures defined via the $H_m$--PVI analysis during our slow-to-moderate solar wind intervals with enhanced $\alpha/p$ and heavy ion charge states. (a) Autocorrelations of the $H_m^{-}(t)$ profiles from Figures~\ref{fig:hcsints} (top) and \ref{fig:psints} (bottom). (b) Temporal waiting time distribution of $\Delta t$ between PVI peaks for the HS (\#8, \#3) and PS (\#7, \#6, \#5+4, \#1) intervals. (c) Spatial waiting time distribution of $\Delta s$ between PVI peaks for the HS and PS intervals. In columns (b), (c), the red curves show power-law fits to the respective waiting time distributions.}
	\label{fig:stats}
\end{figure*}

Given the variation and ``randomness'' of the magnetic field structure(s) and fluctuations within our slow-to-moderate speed, composition-enhanced HS and PS solar wind intervals, statistical methods are required to characterize various properties of the time series \citep[e.g.][]{Zurbuchen2000}. A summation of these analyses are presented in Figure~\ref{fig:stats}.

Figure~\ref{fig:stats}(a) plots the autocorrelation functions $A_{Hm}(\tau)$ of the $H_m^{-}(t)$ time series of the HS intervals (top row) and the PS intervals (bottom row). The average $e$-folding time, $\langle \tau_{1/e} \rangle$, for each set of curves is given in their respective panels. If one defines a characteristic width (duration) of the magnetic helicity-carrying structures as $w = 2 \langle \tau_{1/e} \rangle$ then the mean HS interval width is $w_{\rm HS} = 0.94 \pm 0.02$~hr and the mean PS interval width is  $w_{\rm PS} = 0.96 \pm 0.22$~hr. These values are consistent with, i.e. on the order of, the ${\sim}90$~min periodicity found in solar wind proton density structures \citep[e.g.][]{Viall2010,Viall2015,Kepko2016,DiMatteo2019}.

Figure~\ref{fig:stats}(b) plots the temporal waiting time histogram, $f_{\rm PVI}(\Delta t)$, during the HS (top row) and PS (bottom row) intervals. We have fit a line to each of the distributions in log--log space using the IDL {\tt linfit.pro} least-squares minimization procedure representing a $f(x)=Ax^b$ power law form. The best-fit lines are also plotted in red in each panel and the fit parameters (and their 1-$\sigma$ uncertainties) are given in the plot. The HS and PS distributions have very similar slopes: $b = -0.83 \pm 0.08$ in the HS case and $b = -1.02 \pm 0.06$ for the PS case. If the PVI peaks represent boundaries of coherent magnetic structures, i.e. plasmoid flux ropes or individual flux tubes, then the $\Delta t$ ``waiting time'' between PVI peaks should be roughly the flux structure's diameter (with some variation due to the spacecraft's relative impact parameter). The mean waiting times are $\langle \Delta t \rangle$ = 1.10~hr and 1.01 hr for HS and PS distributions, respectively. The vertical yellow bar in the HS waiting time distribution highlights the bins centered at $\Delta t =$ 1.625, 1.875, and 2.125~hr. Each of these bins having counts $\gtrsim$ 1-$\sigma$ above the best-fit line may indicate the presence of additional coherent structure at these timescales which is, again, remarkably consistent with the \citet{Viall2010} ${\sim}90$~min timescales for periodic density structures. Interestingly, the PS waiting time distribution does not appear to have a similar enhancement in the 1.5--2~hr scale range, although the counts in the PS bins at $\Delta t$= 2.875 and 5.62~hr are also on the order of 1-$\sigma$ above the best fit line.

Figure~\ref{fig:stats}(c) plots the spatial waiting time histogram, $f_{\rm PVI}(\Delta s)$, in the same format as column \ref{fig:stats}(b). Here we note the mean spatial lengths for the HS and PS intervals are, again, essentially identical at $\langle \Delta s \rangle = 2.44 \, R_{\odot}$ ($1698$~Mm) and $2.41 \, R_{\odot}$ ($1677$~Mm), respectively. An interesting feature is the ``disappearance'' of the small enhancement at the 1.5--2~hr scale range in the HS PVI waiting time distribution when plotted as spatial scales. Since we used the radial velocity time series to integrate the distance between PVI peaks, rather than a constant $V_r$ value, the PVI $\Delta s$ distribution is not merely a re-scaled version of the $\Delta t$ distribution. This means, at least in the case of HS slow-to-moderate speed solar wind, that it may be possible to miss a periodic or quasi-periodic signal associated with solar wind formation/source region properties during its subsequent heliospheric evolution if one is focusing on the spatial domain. Conversely, the counts in the PS $\Delta s = 1.25\, R_\odot$ bin are significantly above the power law fit without an obvious corresponding enhancement in the PS $\Delta t$ distribution. The average solar wind speed obtained from the first moment of the temporal and spatial times are $\langle V_r \rangle = \langle \Delta s \rangle/\langle \Delta t \rangle = 429$~km~s$^{-1}$ for the HS intervals and $\langle V_r \rangle = 461$~km~s$^{-1}$ for the PS intervals. These values appear to be slightly lower than the averages obtained directly from the $V_r(t)$ profiles during our composition-enhanced intervals (Table~\ref{tab:int}, Figures~\ref{fig:int}--\ref{fig:backmap}).

Our PVI waiting time statistics seem compatible and consistent with previous applications of these analyses; at scales below the magnetic correlation scale, the PVI waiting time distribution is well approximated by a power-law and at scales greater than the correlation scale, the distribution takes on more of the classic Poisson waiting time exponential form \citep{Greco2009a,Greco2009b}. The temporal/spatial plots in Figure~\ref{fig:stats}(b),(c) show a consistent departure/roll-over from the best-fit line for $\Delta t \gtrsim 2.4$~hr ($\Delta s \gtrsim 6\,R_\odot$) and the first moments of the waiting times/length scales ($\langle \Delta t \rangle$, $\langle \Delta s \rangle$) are on the order of the associated correlation scales (cf. $\S$\ref{sec:results:hmpvi}). In fact, the range of values we obtain for the power-law fit exponents (-1.02 to -0.78) are entirely consistent with those found by \citet{Greco2009b} in MHD turbulence simulation data (-0.92) and in the solar wind at 1~au (-1.29), and in PSP observations of the PVI $> 3$ magnetic field fluctuations at $\sim$0.25~au \citep[-1.29 to -0.83;][]{Chhiber2020}.

\section{Summary and Discussion}
\label{sec:disc}

It is well established that in-situ solar wind composition, its variation and its associated plasma structures are all remnant signatures of the physical processes of solar wind formation and the coronal conditions of its origin. We have presented a comprehensive analysis of a set of slow-to-moderate speed, composition-enhanced solar wind intervals at 1~AU during CR~2002. Our intervals were selected on the basis of solar wind speed and observed enhancements in some combination of $n_p$, $n_\alpha$, $A_{\rm He}$ or their variability. We have shown that each of these intervals correspond to solar wind flows with complex, broadened or bidirectional suprathermal electron strahl, elevated (hot) ionic charge states of C, O, and Fe, and an enhanced Fe/O ratio. 

Pseudostreamers are a prime location for interchange reconnection and they are thought to be responsible for a component of intermittent, slow solar wind outflow \citep[e.g.][]{Masson2012, WangYM2012a, Higginson2017b, WangYM2019a}. In general, energizing surface flows (e.g. translation or rotational shearing flows, flux emergence, and/or flux cancellation/tether-cutting) will build up volumetric currents, stress magnetic null points, and develop strong current sheets at topological boundaries, thereby creating favorable conditions for magnetic reconnection \citep[e.g.][]{Antiochos2012, Rappazzo2012, Burkholder2019, Mason2021}.

\citet{Lynch2013} showed that 2.5D pseudostreamer interchange reconnection (in the form of pre-eruption breakout reconnection) could result in bursty, quasi-steady signatures in density along the external spine and coronal dimming signatures near the stressed null point and current sheet \citep[see also][]{Kumar2021}, while recent simulations from \citet{Aslanyan2021, Aslanyan2022} have illustrated that the complex 3D interchange reconnection dynamics seen by \citet{Higginson2017a,Higginson2017b} can also be produced at the open--closed boundaries of pseudostreamer flux systems.

There is an implicit relationship between the \citet{ZhaoL2017a} source region categories and the large-scale coronal magnetic topology in the neighborhood of the PFSS field line foot points. For example, their `Quiet Sun,' `Active Region,' and `Active Region Boundary' classifications---typically thought of as closed flux regions---are likely to be associated with structures giving rise to the S-Web, i.e. pseudostreamers and small/narrow open field regions such as low-latitude coronal holes. With the application of standard backmapping techniques, we showed that the slow-to-moderate speed, composition-enhanced solar wind intervals at 1~AU map to large-scale coronal features such as the HS belt and S-Web arcs. These are precisely the locations predicted by the $Q$-map topology analysis to be sites favorable for interchange reconnection during the dynamic evolution of the solar corona's open--closed flux boundaries. Lastly, we note that the presence of relatively slow, highly structured, and composition-enhanced solar wind that originates from S-Web arcs far from the HCS is a crucial test of the S-Web theory \citep[e.g.][]{Higginson2017b, DiMatteo2019}.

We have analyzed properties of the in-situ coherent magnetic structures within each composition-enhanced interval as determined by the \citet{Pecora2021} $H_m$--PVI procedure for the identification of helicity-carrying flux tubes and/or magnetic island plasmoids. The characteristic widths of these coherent magnetic structures ($\sim$1 hr from $H_m$; $\sim$2~hr from PVI) are consistent with the ${\sim}90$~min periodicities determined from either in-situ proton density time series \citep{Viall2010} or in the Thomson-scattered white-light coronagraph brightness fluctuations that are proportional to the line-of-sight integrated electron density $n_e$ \citep{Viall2015}. There appears to be a 1.5--2~hr timescale signature above the expected power-law distribution of PVI waiting times in HS-associated solar wind that is either significantly less obvious or non-existent in our PS intervals. There also appears to be an enhancement of the PS-associated waiting time length scale $s \sim 1.25\,R_\odot$ without a corresponding enhancement in the temporal distribution. One may expect different types of reconnection-generated magnetic structures at the boundaries of HS and PS regions due to the topological differences, e.g. as discussed by \citet{Edmondson2017} and \citet{Higginson2018}, but further numerical modeling of their origin and heliospheric evolution will be needed.

This work complements previous statistical studies characterizing magnetic field and plasma properties within coherent intervals or by solar wind type \citep[e.g.][]{Ko2018,DAmicis2019,Borovsky2019,Borovsky2021b}, as well as those studies of specific, small-scale structures \citep{Khabarova2021,Gershkovich2022}, such as small magnetic flux ropes \citep[e.g.][]{Feng2008,Yu2016,MurphyA2020,Choi2021}. Importantly, our attempt to relate various in-situ properties of the structued variability in slow-to-moderate speed solar wind through an application of the $H_m$--PVI methodology represents a significant extension of previous work where coherent magnetic structures identified ``by eye'' were shown to be coincident with structure in the proton density and $A_{\rm He}$ observations \citep[e.g.][]{Kepko2016,DiMatteo2019}. Given recent interest in the further refinement and development of sophisticated automated methods such as machine learning/artificial intelligence neural networks, the $H_m$--PVI procedure appears to be a promising candidate for inclusion in the suite of tools being constructed to classify solar wind types and properties (e.g., as discussed in Section~\ref{sec:intro}) and to identify and characterize coherent flux rope intervals, ranging in spatiotemporal scales from ICMEs \citep[e.g.][]{Nguyen2019, dosSantos2020, Roberts2020, Narock2022} to small-scale flux ropes embedded in the slow solar wind and HCS/HPS crossings \citep[e.g.][]{HuQ2018,ZhaoLL2020}.

The results presented herein open up a number of avenues for future research efforts: (1) extending the current analysis to in-situ solar wind plasma, field, and composition measurements to many more CRs over different phases of the activity cycle; (2) performing forward modeling of heavy ion charge states and elemental abundances associated with the spatial distribution of discrete, observer-connected solar wind flux tubes with varying solar wind outflow properties based on coronal conditions of their foot point locations/source region topologies; and (3) further analysis of existing and future numerical MHD simulations of dynamic S-Web outflow and their derived observational signatures. 

Since there has been recent progress integrating aspects of heavy ion composition forward modeling into steady-state MHD solar wind calculations \citep[e.g.][]{Oran2015, Shen2017, Lionello2019, Szente2022}, it would be extremely interesting to perform these calculations on dynamic, time-dependent MHD modeling of the formation and evolution of coherent magnetic structures generated under different reconnection scenarios. For example, the \citet{Aslanyan2022} calculation of the synthetic suprathermal electron PAD ``time series'' associated with PS interchange reconnection outflows shows excellent qualitative agreement with the observed broadening of the strahl for some of our PS intervals (\#s 4, 6, and 7, in particular). \citet{Lynch2014} showed the largest (i.e. low frequency) $\delta B/\langle B \rangle$ signatures resulting from PS reconnection had characteristic length-scales of 100--350~Mm (0.14--0.50 $R_\odot$) in the corona which reflected the spatial scale of their PS flux system of origin, and \citet{Higginson2018} demonstrated that the MHD simulation-derived, synthetic in-situ magnetic field signatures of a similarly-sized, non-linear torsional Alfv\'{e}n wave could resemble the coherent magnetic structure of small-scale magnetic flux ropes/streamer blob plasmoids typically associated with HS slow wind in the vicinity of the HCS/HPS.

On the largest scales (the 10s of hours of our interval durations), there is a remarkably clear association between our HS and PS S-Web arc intervals and in-situ composition enhancements. On the scales of coherent magnetic structures depicted in Figures~\ref{fig:hcsints}--\ref{fig:psints}, there are \emph{some} indications that the PVI boundaries are also associated with discrete changes in coronal freeze-in temperatures as inferred from the heavy ion charge states. The 2-hr cadence of the ACE/SWICS data used herein obviously limits our ability to resolve charge-state structure below the averaging window duration. Smaller scale features have been observed and reported in \citet{Kepko2016} and \citet{Gershkovich2022} using periods of high-cadence (12-min native instrument resolution) ACE/SWICS data to argue that some of the discrete magnetic flux tube intervals of interest did line up with sudden changes in various composition measures (i.e. He, C, and O abundances, the C$^{6+}$/C$^{5+}$ charge state ratio, etc). Measurements from the Heavy Ion Sensor (HIS, with a native resolution of 30 seconds for heavy ions), part of the \emph{Solar Orbiter} Solar Wind Analyser (SWA) instrument suite \citep{Owen2020}, should enable the identification and characterization of smaller scale associations of coherent magnetic structures with in-situ composition enhancements.  

Additionally, the scientific importance for multispacecraft measurements and remote-sensing and in-situ quadrature observational geometries, to establish the solar--heliospheric connection for specific plasma features of well-observed interchange reconnection events has been recently demonstrated by \citet{Telloni2022} with the first direct imaging of a ``switchback'' with \emph{Solar Orbiter}'s Metis coronagraph \citep{Antonucci2020}. This switchback event's likely origin from the complex S-Web configuration of a small PS S-Web arc coming off the main HS belt/HCS and a null-point spine--fan curtain topology where the PS and HS flux systems intersect strongly motivates continued theoretical development, data analysis, and numerical simulations of the dynamic S-Web model for the slow solar wind.  

\acknowledgments

The authors acknowledge support from NASA HGI 80NSSC18K0645 and HSR 80NSSC18K1553. Additionally, BJL acknowledges support from NASA grants 80NSSC21K0731, 80NSSC20K1448, and 80NSSC21K1325; NMV and AKH acknowledge support from the competed Internal Scientist Funding Model (IFSM) at NASA GSFC; LZ acknowledges support from NASA grant 80NSSC21K0579; and STL acknowledges support from NASA grants 80NSSC19K0853 and 80NSSC20K0192.

The authors thank the \emph{Wind} and ACE mission teams for making the in-situ magnetic field, plasma, and composition data available at \url{http://cdaweb.gsfc.nasa.gov} and \url{https://izw1.caltech.edu/ACE/ASC/}, as well as the MDI team for making the photospheric magnetogram data available at \url{http://hmi.stanford.edu/data/synoptic.html}.

%



\bibliographystyle{apj} 
\bibliography{apj-jour,master}

\end{document}